\documentclass[twocolumn,superscriptaddress,nofootinbib,notitlepage,longbibliography,aps]{revtex4-1}
\pdfoutput=1

\usepackage{graphicx}
\usepackage{amsmath}
\usepackage{amssymb}
\usepackage{amsfonts}
\usepackage{physics}
\usepackage[dvipsnames]{xcolor}
\usepackage[linktoc=none]{hyperref}
\usepackage{braket}
\usepackage{tabstackengine}
\stackMath
\usepackage[mathscr]{euscript}

\hypersetup{colorlinks,linkcolor={blue},citecolor={teal},urlcolor={violet}}

\newcommand{\INFN}{INFN - Sezione di Napoli, Complesso Universitario Monte S. Angelo, I-80126 Napoli, Italy}
\newcommand{\UNINA}{Dipartimento di Fisica ``Ettore Pancini'', Università degli studi di Napoli ``Federico II'', Complesso Universitario Monte S. Angelo, I-80126 Napoli, Italy}
\newcommand{\SSM}{Scuola Superiore Meridionale, Via Mezzocannone 4, 80138 Napoli, Italy}

\newcommand{\MPBH}{{M}_{\rm PBH}}

\newcommand{\TPBH}{\mathcal{T}_{\rm PBH}}

\newcommand{\GammaNth}{\Gamma_{N_1}^{\rm th.}}
\newcommand{\GammaNPBH}{\Gamma_{N_1}^{\rm PBH}}
\newcommand{\GammaPBHtoN}{\Gamma_{{\rm PBH}\to N_1}}

\newcommand{\YB}{{Y}_{\rm B}}

\newcommand{\CED}{\mathcal{S}} 

\newcommand{\YPBH}{\mathcal{N}_{\rm PBH}}
\newcommand{\YNPBH}{\mathcal{N}_{N_i}^{\rm PBH}}
\newcommand{\YNth}{\mathcal{N}_{N_1}}
\newcommand{\Yeq}{\mathcal{N}_{N_1}^{\rm eq.}}
\newcommand{\Yeqlep}{\mathcal{N}_{\ell}^{\rm eq.}}
\newcommand{\YBL}{\mathcal{N}_{\rm B-L}}

\newcommand{\varrhopbh}{\varrho_{\rm PBH}}
\newcommand{\varrhorad}{\varrho_{\rm rad}}

\newcommand{\rhopbh}{\rho_{\rm PBH}}

\newcommand{\rhorad}{\rho_{\rm rad}}

\newcommand{\Tform}{T_{\rm in}}

\newcommand{\Mpl}{{M}_{\rm Pl}}

\newcommand{\rhocr}{\rho_{\rm cr.}}

\begin{document}

\title{Impact of memory-burdened primordial black holes on high-scale leptogenesis}

\author{Roberta Calabrese}
\email{rcalabrese@na.infn.it}
\affiliation{\UNINA}
\affiliation{\INFN}

\author{Marco Chianese}
\email{m.chianese@ssmeridionale.it}
\affiliation{\SSM}
\affiliation{\INFN}

\author{Ninetta Saviano}
\email{nsaviano@na.infn.it}
\affiliation{\INFN}
\affiliation{\SSM}

\begin{abstract}
We explore the impact of the backreaction of evaporation on the quantum state of primordial black holes (PBHs), known as ``memory burden", on the baryon asymmetry production in the Universe through high-scale leptogenesis. Focusing on PBH masses ranging from 1 to 1000 grams, we investigate the interplay between the nonthermal production of heavy sterile neutrinos and the entropy injection within this nonstandard cosmological framework. By assuming appropriate values for the memory-burden parameters, $q=1/2$ and $k=1$, we derive mutual exclusion limits between PBHs and thermal leptogenesis in the mixed parameter space. Our analysis reveals that the primary contribution of PBHs to baryon asymmetry stems from entropy injection. Indeed, we find that, differently from earlier studies based on the semiclassical Hawking evaporation, the memory-burden effect suppresses the nonthermal source term in the PBH mass range explored. This has significant implications for understanding baryogenesis in such alternative cosmological scenarios.
\end{abstract}
\maketitle

\section{Introduction}
The  seminal idea that Primordial Black Holes (PBHs) have been formed in the  very early universe from dense regions  long before stars emerged \cite{Zeldovich:1967lct, Hawking:1971ei, Carr:1974nx}, has been recently resumed, gaining new interest in relation to the context of dark matter~\cite{Bernal:2020kse, Gondolo:2020uqv, Bernal:2020ili, Bernal:2020bjf, Cheek:2021odj, Cheek:2021cfe, Samanta:2021mdm, Bernal:2021yyb, Bernal:2021bbv, Sandick:2021gew, Bernal:2022oha, Cheek:2022mmy, Gehrman:2023qjn, Bertuzzo:2024fns, Basumatary:2024uwo, Basumatary:2024uwo, Domenech:2024wao}, early Universe investigation~\cite{Carr:2016drx, Green:2020jor, Carr:2020gox, Carr:2021bzv}, as well as gravitational wave research~\cite{Papanikolaou:2020qtd, Domenech:2020ssp, Papanikolaou:2022chm, Ireland:2023avg}. The mass and spin distributions of PBHs can vary based on the specific cosmological scenario. PBHs with masses above 
$10^{15}$ grams are stable over cosmological timescales, but lighter black holes may have already evaporated through Hawking radiation by the current era. This process of evaporation has been investigated in a variety of different contexts, playing  important roles in the production of the baryon asymmetry of the Universe, gravitational wave emission, or dark matter generation, depending on their mass. Black hole evaporation, as described by the semiclassical framework, assumes that the black hole retains its classical nature throughout its entire lifetime~\cite{Hawking:1975vcx}. However, it is becoming increasingly evident that this assumption may lead to an inconsistent model, pointing to the necessity for new physics, especially in relation to the information loss paradox~\cite{Preskill:1992tc}. Hawking's analysis fails to account for the backreaction of emitted particles on the quantum state of the black hole. This omission becomes particularly important when the energy of the emitted quanta approaches the black hole’s total energy. Recent research~\cite{Dvali:2018xpy, Dvali:2024hsb} suggests that such backreaction could give rise to a ``memory burden'', a phenomenon where the system's information resists being lost, driven by the response of quantum modes linked to its entropic degrees of freedom. As the black hole's mass decreases past a certain point, the backreaction becomes increasingly significant, potentially slowing the evaporation rate and extending the black hole's lifetime. Among the significant phenomenological consequences \cite{Balaji:2024hpu, Barman:2024iht, Bhaumik:2024qzd, Barman:2024ufm, Kohri:2024qpd, Jiang:2024aju, Dvali:2020wft, Dvali:2021byy, Alexandre:2024nuo, Thoss:2024hsr, Haque:2024eyh, Chianese:2024rsn, Zantedeschi:2024ram, Barman:2024kfj, Borah:2024bcr, Athron:2024fcj, Loc:2024qbz, Bandyopadhyay:2025ast}, the potential of the memory burden could also impact the  baryogenesis via leptogenesis scenarios~\cite{fukugita1986barygenesis}, where the lepton asymmetry, generated from lepton number violating interactions, is converted to a net excess of baryons via the non-perturbative $(B+L)$-violating sphaleron transitions \cite{Davidson:2008bu, Fong:2012buy,Buchmuller:2005eh,DiBari:2012fz,Harvey:1990qw,Kuzmin:1985mm,Khlebnikov:1988sr}. Normally, depending on the mass and so on the temperature of the PBHs, the evaporation of primordial black holes in the early Universe could have impacted the process of leptogenesis  through both the production of heavy right-handed neutrinos, and the injection of entropy in the plasma, with production of significant population of relativistic particles~\cite{Perez-Gonzalez:2020vnz, JyotiDas:2021shi, Bernal:2022pue, Calabrese:2023key, Calabrese:2023bxz, Gunn:2024xaq}. The presence of the memory-burden effect has the potential to alter this scenario, as it could extend the lifespan of PBHs. Consequently, considering certain values of their abundance, PBHs could come to dominate the Universe during a specific period when, under normal circumstances in the semiclassical regime, they would not have been the dominant component. In this study of thermal leptogenesis, we find that the primary influence of memory-burdened PBHs lies not in their role as a nonthermal source of heavy sterile neutrinos, which could otherwise enhance the baryon asymmetry production, but rather in their entropy injection, leading to a net reduction in the baryon asymmetry. 
We find that the memory-burden effect notably alters the parameter space, shifting the viable region toward lower PBH masses in comparison to previous studies.

The paper is organized as follows. In Sec.~\ref{sec:PBH_cosmology}, we describe the evaporation process of memory-burden PBHs and the corresponding impact on the cosmological history of the Universe. In Sec.~\ref{sec:HSL}, we detail the particle model of the thermal high-scale leptogenesis. In Sec.~\ref{sec:boltz}, we discuss the Boltzmann equations dictating the evolution of the baryon asymmetry of the Universe. In Sec.~\ref{sec:res}, we report the main results of the present analysis. Finally, in Sec.~\ref{sec:concl} we draw our conclusions.

\section{Cosmology with Memory-Burdened Primordial Black Holes}\label{sec:PBH_cosmology}

In this Section, we describe how the existence of memory-burdened PBHs can influence the evolution of the Early Universe. Thanks to the memory-burden effect, PBHs live longer allowing their domination for smaller masses compared to the standard scenario. 
For the sake of concreteness, we consider neutral and non-rotating PBHs with a monochromatic mass distribution with initial mass $\MPBH^{\rm in}$. Moreover, we assume they originate from the collapse of density perturbations when they reenter the horizon in a radiation-dominated Universe at the temperature
\begin{equation}
    \Tform=\frac{1}{2}\,\left(\frac{5}{\pi^3 g_*}\right)^{1/4}\sqrt{\frac{3\gamma_{\rm PBH} \Mpl^3}{\MPBH^{\rm in}}}\,,
    \label{eq:T_in}
\end{equation}
where $\gamma_{\rm PBH}\sim0.2$ is the gravitational collapse factor~\cite{Carr:1975qj, Carr:2016drx}, and $\Mpl = 1.22\times10^{19}$\,GeV is the Planck mass. We refer to the initial PBHs abundance in terms of the quantity
\begin{equation}
    \beta^\prime = \gamma^{1/2}_{\rm PBH}\left(\frac{g_*(\Tform)}{106.5}\right)^{-1/4}\left(\frac{h}{0.67}\right)^{-2}\left. \frac{\rhopbh}{\rhocr}\right|_{\rm in}\,,
    \label{eq:beta_prime}
\end{equation}
where $h = 0.67$\,\cite{Planck:2018jri} is related to the Hubble constant, and $\rhopbh$ and $\left.\rhocr \right|_{\rm in } = \frac{3\gamma_{\rm PBH}^2}{32\pi}\frac{\Mpl^6}{(\MPBH^{\rm in})^2}$ are the PBH and critical energy densities at $\Tform$, respectively. 

As soon as PBHs are formed, they radiate all the particles with a mass smaller their Hawking temperature $\TPBH$, defined as~\cite{Hawking:1975vcx, Hawking:1974rv}
\begin{equation}
    \TPBH \simeq 10^{10}    \left(\frac{10^{3}~{\rm g}} {\MPBH} \right){\rm GeV}\,.
\label{PBH temperature}
\end{equation}
Owing to the memory-burden effect, the evaporation process occurs into two distinct phases. The first phase is the semiclassical (Hawking-like) one, where the PBH mass evolves according to
\begin{equation}
\frac{{\rm d}\ln \MPBH}{{\rm d}\alpha} = - \frac{\kappa(\MPBH)}{H \MPBH^3}\,,\label{eq:M_evol_pt1}
\end{equation}
where $\alpha=\ln a$ with $a$ being the cosmic scale factor, and $\kappa(\MPBH) = 416.3\,\Mpl^4/(30720\pi)$ is constant for $\MPBH\lesssim 10^{9}$\,g~\cite{Datta:2020bht}. The second ``memory-burdened phase'', where the PBHs are stabilized by the memory burden, kicks in when $\MPBH = q \MPBH^{\rm in}$. In the second phase, quantum effects become relevant: the information stored on the event horizon back-reacts and slows down the evaporation process by a factor proportional to a negative power of the PBH entropy
\begin{equation}
S_{\rm PBH} = 4\pi G \MPBH^2\,.
\end{equation}
Hence, the mass evaporates according to 
\begin{equation}
\frac{{\rm d}\ln \MPBH}{{\rm d}\alpha} = - \frac{1}{S_{\rm PBH}^k}\frac{\kappa(\MPBH)}{H \MPBH^3}\,,\label{eq:M_evol_pt2}
\end{equation}
Taking into account both the two evaporation phases, the number of $\chi$ particles emitted by a single PBH is
\begin{eqnarray}
\frac{\dd N_\chi}{\dd E\dd t} &=& \frac{g_\chi}{2\pi} \frac{\mathcal{F}_\chi(E, \MPBH)}{e^{E/\TPBH} + (-1)^{2s_\chi}}\nonumber \\ &&\,\,\times \left\{\begin{array}{cc}1 & \MPBH>q\MPBH^{\rm in} \\ {S^{-k}_{\rm PBH}}& \text{otherwise}\end{array}\right.\,.
\end{eqnarray}
where $g_\chi$ and $s_\chi$ are the degrees of freedom and the spin of $\chi$ particles, respectively, and $\mathcal{F}_\chi$ is the grey-body factor\,\cite{Perez-Gonzalez:2020vnz}. By integrating this equation over the energy, we obtain the emission rate 
\begin{equation}
\Gamma_{{\rm PBH}\to \chi} = \int_{m_\chi}^\infty \frac{\dd N_\chi}{\dd E\dd t}\dd E \,.
\label{EQ:gamma_PBH_to_chi}
\end{equation}
The parameter space of the memory-burden effect is therefore completely determined by the two additional parameters $q$ and $k \geq 0$. In the present analysis, we fix $q=1/2$ and $k=1$ and leaves the study of different memory-burden parameters to future works. We note that such choice guarantees that PBHs with mass $\MPBH^{\rm in} \leq 1000~{\rm g}$ have been completely evaporated before the onset of Big Bang Nucleosynthesis~\cite{Haque:2024eyh}.

To follow the evolution of the main energy density components, we need to solve the following coupled Friedmann equations~\cite{Lunardini:2019zob, Perez-Gonzalez:2020vnz, Bernal:2022pue}
\begin{eqnarray}
\frac{{\rm d}\varrhorad}{{\rm d}\alpha} &=& - f_{\rm SM} e^{\alpha} \frac{{\rm d}\ln \MPBH}{{\rm d}\alpha}  \varrhopbh\,, \label{eq:FE_rad}\nonumber\\
    \frac{{\rm d}\varrhopbh}{{\rm d}\alpha} &=& \frac{{\rm d}\ln \MPBH}{{\rm d}\alpha}\varrhopbh\,,
\label{eq:FE_PBH}\\
    H^2 &=& \frac{8\pi}{3\Mpl^2}\left(\frac{\varrhopbh}{a^3}+\frac{\varrhorad}{a^4} \right)\,\,,\label{H}\nonumber
\end{eqnarray}
where $H$ is the Hubble parameter, $\varrhopbh \equiv a^3 \rhopbh$ and $\varrhorad \equiv a^4 \rhorad$ are the comoving energy densities of PBHs and radiation, respectively, and $f_{\rm SM}$ is the fraction of Hawking radiation composed of SM particles. In our framework, $f_{\rm SM} \simeq 1$ since PBHs produce a negligible amount of right-handed neutrinos. Notice that we recover the standard evolution for $\varrhopbh = 0$. The comoving entropy density of the Universe evolves as 
\begin{equation}
    \frac{{\rm d}\CED}{{\rm d}\alpha} = -\frac{f_{\rm SM}}{T(\alpha)} \frac{{\rm d}\ln \MPBH}{{\rm d}\alpha} \varrhopbh\,,
    \label{eq:entropy}
\end{equation}
which is constant for $\varrhopbh \to 0$. Lastly, the Universe's temperature $T$ evolves according to 
\begin{equation}
    \frac{\dd T}{\dd\alpha} \simeq -T \left[1 - \frac14 f_{\rm SM} e^\alpha \frac{\dd \ln \MPBH}{\dd\alpha}\frac{\varrhopbh}{\varrhorad}\right]\,.
    \label{eq:T_univ}
\end{equation}

\section{High-scale leptogenesis}\label{sec:HSL}

We now describe in detail the high-scale thermal leptogenesis scenario that lays the foundation of our analysis. Thermal leptogenesis relies on a minimal extension of the Standard Model (SM) incorporating at least two singlet Majorana neutrinos. These neutrinos interact with the SM lepton and Higgs doublets, introducing dynamics that facilitate the generation of a matter-antimatter asymmetry. A particularly compelling feature of this minimal framework is its ability to simultaneously account for the experimentally observed neutrino masses and the baryon asymmetry of the Universe.

\subsection*{Type-I seesaw}

The small masses of light, active neutrinos can be naturally explained via the type-I seesaw mechanism, characterized by the following Lagrangian terms
\begin{equation}
    \mathcal{L} \supset - Y_{\alpha i} \overline{L}_\alpha  \tilde{\phi} N_i  -\frac{1}{2} \overline{N^c}_i \hat{M}_{ij} N_j + {\rm h.c.}\,,
\label{Lagrangian}
\end{equation}
where $\tilde{\phi} = i \sigma_2\phi^*$, $\phi$ is the Higgs doublet, $L_\alpha = (\nu_\alpha,\ell_\alpha)$ are the lepton doublets where $\alpha=e,\mu,\tau $, and $N_i$, $i=1,2,3$ are the Majorana right-handed neutrinos. 

The active neutrino mass matrix is then determined by the type-I seesaw relation\,\cite{Yanagida:1980xy,MohapatraRabindraSenjanovi}
\begin{equation}\label{TI}
m_\nu\simeq -v_{\rm EW}^2\, Y\cdot\frac{1}{\hat{M}}\cdot Y^T \,,
\end{equation}
where $v_{\rm EW}=174$\,GeV is the Higgs vacuum expectation value, $\hat{M}$ represents the mass matrix of the right-handed neutrinos, and $Y$ is the Yukawa coupling matrix from Eq.~\eqref{Lagrangian}. We consider the mass eigenbasis of the right-handed neutrinos so that $\hat{M} = \mathrm{diag}(M_1, M_2, M_3)$. The $3\times 3$ matrix $m_\nu$ can be diagonalized by the lepton mixing matrix $U_{\rm PMNS}$~\cite{pontecorvo1957mesonium, pontecorvo1957inverse, Maki:1962mu, Pontecorvo:1967fh, Gribov:1968kq}, parametrized by three mixing angles, together with one Dirac and two Majorana phases.

We recast Eq.~\eqref{TI} in terms of the physical quantities~\cite{Casas:2001sr}
\begin{equation}\label{Y}
Y = \frac{1}{v_{\rm EW}}\sqrt{\hat{M}}\cdot R \cdot \sqrt{\hat{m}_{\nu}} \cdot U^{\dagger}_{\rm PMNS} \,,
\end{equation}
where $\hat{m}_\nu = \mathrm{diag}(m_1, m_2,  m_3)$ is the diagonalized left-handed neutrino mass matrix, and $R$ is the Casas-Ibarra orthogonal matrix parametrized by three complex angles $z_{12},z_{13},z_{23}$. 
While solar and atmospheric neutrino oscillation experiments have provided precise measurements of the PMNS mixing angles and the two mass-squared differences ($\Delta m_{\rm sol}^2$ and $\Delta m_{\rm atm}^2$), the absolute neutrino mass scale and the mass ordering remain unknown. The ordering may be either normal ($m_1<m_3$) or inverted ($m_1>m_3$)~\cite{Capozzi:2021fjo, Esteban:2020cvm, deSalas:2020pgw}. In the following discussion, we focus on the normal mass ordering\,\cite{Capozzi:2021fjo}, where the heaviest ($m_h$) and the lightest ($m_l$) active neutrino masses correspond to $m_3$ and $m_1$, respectively. Additionally, we adopt a minimal model of thermal leptogenesis with a hierarchical mass spectrum for the right-handed neutrinos, such that $M_1\ll M_2,\, M_3$. As shown in Ref.~\cite{Hambye:2003rt}, being $m_1 \simeq m_2$ because of $\Delta m_{\rm sol}^2 \ll \Delta m_{\rm atm}^2$, the only relevant angle in the $R$ matrix in Eq.~\eqref{Y} is $z_{13}=x+i\,y$ whit $x,y$ being real parameters. This reduces the $R$ matrix to 
\begin{equation}
    R = \begin{pmatrix}
     \cos z_{13} &  0 & \sin z_{13} \\
     0 & 1 & 0\\
     -\sin z_{13} & 0 & \cos z_{13} \\
    \end{pmatrix}\,.
    \label{Rmatrix_simple}
\end{equation}

\subsection*{CP asymmetry}

CP violation in the lepton sector is pivotal in generating the observed baryon asymmetry of the Universe through leptogenesis. Specifically, a non-zero CP asymmetry arises in the out-of-equilibrium decays of heavy right-handed Majorana neutrinos. In our scenario, the dominant contribution to the baryon asymmetry originates from the decay of $N_1$ into SM leptons and Higgs ($N_1\to \ell_\alpha \, \phi$), provided that this decay process is CP asymmetric at the one-loop level~\cite{Davidson:2008bu}. 
The CP asymmetry for $N_1$ decaying into lepton flavor $\alpha$ is 
\begin{equation}
\epsilon_{\alpha\alpha}=\frac{\Gamma(N_1\to \ell_\alpha \, \phi)-\Gamma(N_1\to \overline{\ell}_\alpha\, \overline{\phi})}{\Gamma(N_1\to \ell_\alpha \,\phi)+\Gamma(N_1\to \overline{\ell}_\alpha\, \overline{\phi})}\,,
\end{equation}
where $\Gamma$ indicates the decay width. At temperatures $T \gtrsim 10^{12}$\,GeV, the charged lepton Yukawa interactions are slower than the Hubble expansion rate, rendering leptogenesis insensitive to individual lepton flavors~\cite{Davidson:2008bu}. Since we mainly focus on $M_1 \gtrsim 10^{11}$\,GeV, we can neglect flavor effects and sum over the lepton flavors $\alpha = e,\mu,\tau$. Hence, in the hierarchical limit $(M_1\ll M_{2,3})$, the total CP asymmetry is~\cite{Strumia:2006qk}
\begin{equation}\label{eps}
\epsilon=\sum_\alpha \epsilon_{\alpha\alpha} = - \frac{3}{16\pi}\sum_{j=2,3} \frac{M_1}{M_j}\frac{\mathrm{Im}(Y Y^\dagger)_{1j}}{(Y Y^\dagger)_{11}} \,.
\end{equation}
In this expression, the leptonic mixing matrix $U_{\rm PMNS}$ drops out, indicating no direct link between the low-scale Dirac CP violation and high-scale CP asymmetry. Using the simplification in Eq.~\eqref{Rmatrix_simple}, the CP asymmetry parameter $\epsilon$ can be expressed in terms of only four unknown parameters $\{x,y, m_h, M_1\}$~\cite{Bernal:2022pue}
\begin{equation}
    |\epsilon| = \frac{3M_1}{16\pi v_{EW}^2}\frac{|\Delta m^2_{\rm atm}|}{m_h + m_l}\frac{|\sin(2x)\sinh(2y)|}{\cosh(2y) - f\cos(2x)}\,,
\end{equation}
where $f \equiv (m_h - m_l)/(m_h + m_l)$ with $m_h = m_3$ and $m_l = m_1 \approx m_2$. The parameter $\epsilon$ governs the CP violation in $N_1$ decays, while thermal production of $N_1$ violates CP with the same magnitude but opposite sign. In this study, we fix $|x| = \pi/4$ and $y = 044$ which are a good approximation to maximize the baryon asymmetry produced~\cite{Bernal:2022pue, Calabrese:2023key}. This choice allows us to make a direct comparison between the scenario of memory-burden PBHs and the standard one explored in previous studies~\cite{Bernal:2022pue}.

When the initial abundance of $N_1$ is negligible, an initial ``anti-asymmetry" forms in the lepton sector. As $N_1$ reaches the thermal equilibrium, it decays when the Universe cools to $T=M_1$, provided that the decay rate exceeds the Hubble expansion rate. This condition is quantified as 
~\cite{Strumia:2006qk,Davidson:2008bu, Buchmuller:2004nz}
\begin{equation}\label{K}
    K\equiv \frac{\tilde{m}_1}{m^*} > 1,
\end{equation}
where
\begin{equation}\label{meff}
    m^* = \frac{8\pi v_{\rm EW}^2}{M_1^2} H\left(T=M_1\right)\,,
\end{equation}
and
\begin{equation}
    \tilde{m}_i = \sum_\alpha \frac{|Y_{\alpha\,i}|^2v_{\rm EW}^2}{M_i}\,,
\end{equation}
characterize the Hubble expansion rate at $T=M_1$ and the decay rate, respectively.

If $N_1$ decays occur out of equilibrium, the asymmetry produced by its decay would cancel the initial anti-asymmetry unless suppressed by ``washout" processes. These processes include inverse decays ($\ell \phi^\dagger \to N_1$ and $\bar{\ell} \phi \to N_1 $) and $2\to2$ scatterings ($\ell\phi^\dagger\to\overline{\ell}\phi$) mediated by off-shell $N_1$. For $K>1$, the washout processes efficiently suppress the anti-asymmetry, allowing a net asymmetry to survive, which defines the ``strong-washout" regime. Conversely, in the ``weak-washout" regime ($K<1$), $N_1$ decays occur later when washout processes are weak, causing the asymmetries to largely cancel unless an additional mechanism generates the $N_1$ population~\cite{Davidson:2008bu}.
Experimental constraints on neutrino masses favor the strong-washout regime, as the weak-washout regime requires $m_h$ to be extremely close to $\sqrt{\Delta m^2_{\rm atm}}$. Therefore, in this analysis, we focus on the strong-washout regime. This also ensures that any contribution from $N_2$ and $N_3$ is negligible, as they are washed out at $T\ll M_{1}$.

\section{Boltzmann equations}\label{sec:boltz}

To determine the baryon asymmetry of the Universe, we solve the Boltzmann equations for the evolution of $N_1$ and $B-L$ incorporating the impact of memory-burdened PBHs. Focusing on the $B-L$ asymmetry is particularly advantageous, as it is conserved during sphaleron processes, which subsequently convert a portion of the lepton asymmetry into baryon asymmetry. Our analysis accounts for the influence of PBHs, which contribute in three significant ways: (1) they can serve as a nonthermal source of $N_1$, (2) they modify the evolution of the Hubble parameter, and (3) they inject entropy at the end of their evaporation, diluting the final asymmetry.

In our analysis, we assume an initial abundance of $N_1$ and $B-L$ asymmetry to be zero. The evolution is governed by the following processes:
\begin{itemize}
\item {\bf Decays ($1\to 2$):} The decay of $N_1$ into leptons and Higgs bosons ($N_1 \to \bar{\ell} \phi$) and their CP-conjugate processes ($N_1 \to \bar{\ell} \phi$). The total decay rates are proportional to the number density of $N_1$ and the thermally averaged decay width. These decays deplete the $N_1$ population while generating a source term for the $B-L$ asymmetry.
\item {\bf Inverse Decays ($2 \to 1$):} Processes such as $\ell \phi^\dagger \to N_1$ replenish the $N_1$ population but act solely as a washout mechanism for the $B-L$ asymmetry. The inverse decay rate is proportional to the equilibrium number densities of $N_1$ and leptons and is related to the decay rate via $\Gamma^{ID}_{N_1} = \Gamma_{N_1}n_{N_1}^{\rm eq}/n^{\rm eq}_{\ell}$, where $n_{N_1}^{\rm eq},n^{\rm eq}_{\ell}$ are their equilibrium number densities.
\item {\bf Scattering Processes ($2 \leftrightarrow 2$):} These include $\Delta L = 2$ scatterings mediated by $N_1$ exchange, such as $\ell \phi^\dagger \to \bar{\ell} \phi$. While these processes do not alter the number density of $N_1$, they contribute to the washout of the $B-L$ asymmetry. To avoid double counting with the s-channel contributions of inverse decays, we use a subtracted form of the scattering amplitude that includes only the off-shell $N_1$ exchange. This ensures these processes remain effective at low temperatures and are not Boltzmann suppressed.
\end{itemize}
Although $2 \leftrightarrow 2$ scatterings involving gauge bosons and top quarks also contribute, their impact is proportional to higher-order couplings $\propto(Y^2 Y_t^2)$, where $Y_t$ is the top quark Yukawa coupling, and are therefore neglected in this analysis. Similarly, three-body decay processes are numerically insignificant, contributing only around 6\%, and are also omitted~\cite{Nardi:2007jp}.
\begin{figure*}[tbh!]
    \centering
    \includegraphics[width=\textwidth]{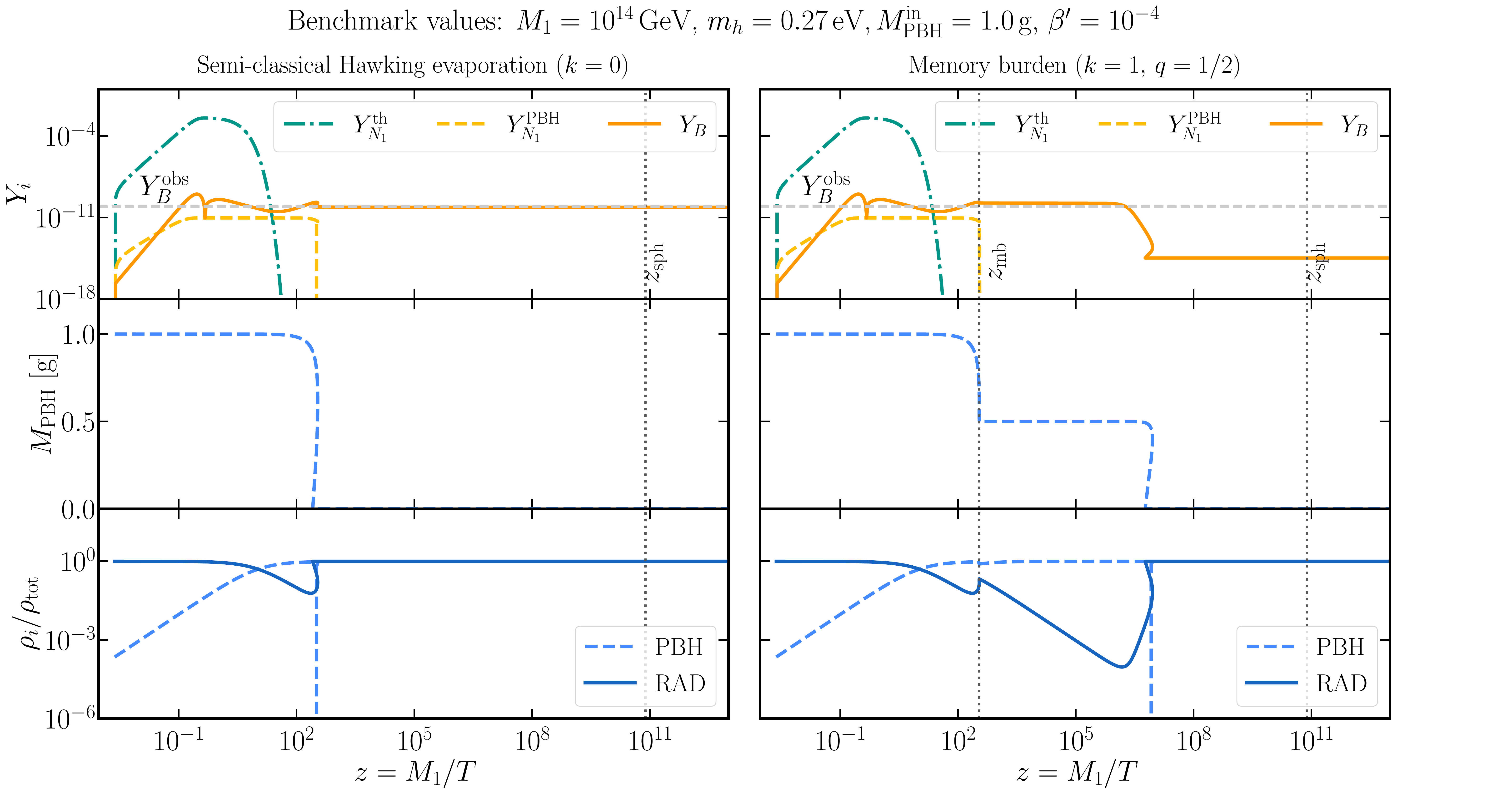}
    \caption{Evolution of the main quantities as a function of the auxiliary variable $z=M_1/T$ for the benchmark case $M_1 = 10^{14}$ GeV, $m_h = 0.27$\,eV, $\MPBH^{\rm in} = 1.0$\,g and $\beta^{\prime} = 10^{-4}$, in case of the semiclassical Hawking evaporation scenario (left plot) and of the memory-burden effect one (right plot). The upper panels illustrate the evolution of the thermal and nonthermal yields of the right-handed neutrinos, shown by green dot-dashed and yellow dashed lines, respectively, alongside the absolute value of the $B-L$ yield (solid orange lines). The horizontal dashed lines correspond to the observed baryon asymmetry. The middle panels display the evolution of the PBH mass, while the lower panels depict the radiation and PBH energy densities normalized by the total energy density. The vertical dotted lines correspond (from left to right) to the onset temperatures of the memory-burden phase and the sphaleron processes.}
    \label{fig:benchmarks_1}
\end{figure*}

\begin{figure*}[tbh!]
    \centering
    \includegraphics[width=0.99\linewidth]{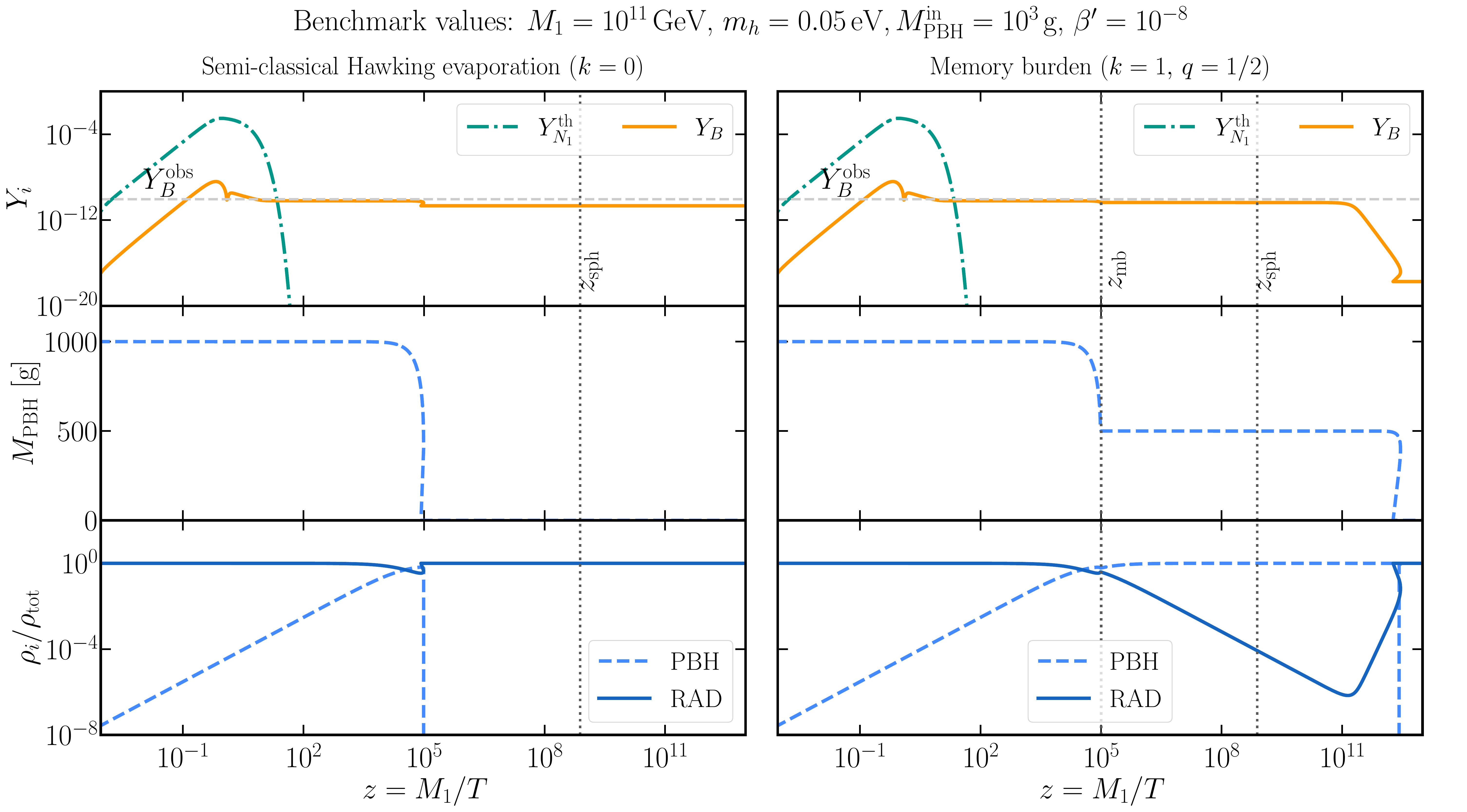}
    \caption{Similar to Fig.~\ref{fig:benchmarks_1}, but assuming $M_1 = 10^{11}$\,GeV, $m_h = 0.05$\,eV, $\MPBH^{\rm in} = 10^3$\,g, and $\beta^\prime = 10^{-8}$.}
    \label{fig:benchmarks_2}
\end{figure*}

Hence, we need to solve the Boltzmann equations for the thermal, nonthermal (from PBHs evaporation) $N_1$ and $B$-$L$ comoving number densities. These equations read as 
\begin{eqnarray}
\frac{\dd\YNth}{\dd \alpha} &=&  \frac{\GammaNth}{H}(\Yeq-\YNth)\,, \label{eq:Nth} \\
\frac{\dd \YNPBH}{\dd \alpha}  &=&\frac{1}{H}\left(\GammaPBHtoN\YPBH +\right. \nonumber \\
&& \left.\qquad \quad - \GammaNPBH \YNPBH \right)\,,\label{eq:non_thermal} \\
\frac{\dd \YBL} {\dd\alpha} &=&   \left.\frac{1}{H}\right[ \epsilon(\YNth-\Yeq)\GammaNth +\nonumber\\
&&\qquad \quad + \GammaNPBH \YNPBH + \\ \nonumber
&&\qquad \quad \left. + \left(\frac{1}{2} \frac{\Yeq}{\Yeqlep}\GammaNth + \gamma \frac{a^3}{\Yeqlep}\right)  \YBL\right]\,,\label{AsymBoltzmann}
\end{eqnarray}
where $\YNth = n_{N_1}a^3$, $\YPBH = \varrhopbh/\MPBH$, and $\YBL = n_{\rm B-L}a^3$.
The quantity $\GammaNth$ is the thermally averaged, tree-level decay rate of $N_1$, given by 
\begin{equation}
    \GammaNth = \left<\frac{M_{1}}{E_{N_1}}\right>  \frac{(Y^\dagger Y)_{11}M_1}{8\pi}\,,
\end{equation}
where the thermal averaging factor $\left<M_1/E_{N_1}\right>$ can be well approximated by the ratio $K_1(z)/K_2(z)$, with $K_n$ being the $n$-th order modified Bessel function of the second kind. The quantity $\GammaNPBH$ in Eq.\,\eqref{eq:non_thermal} is the decay width corrected by an average inverse dilatation factor 
\begin{equation}
\GammaNPBH \approx \frac{K_1(z_{\rm PBH})}{K_2(z_{\rm PBH})} \frac{(Y^\dagger Y)_{11}M_1}{8\pi}\,,
\end{equation}
where $z_{\rm PBH} = M_1/\TPBH$. The PBH production rate $\GammaPBHtoN$ of right-handed neutrinos is reported in Eq.~\eqref{EQ:gamma_PBH_to_chi}. The parameter $\gamma$ in Eq.~\eqref{AsymBoltzmann} quantifies the contribution to the washout from $\Delta L = 2$ scattering processes. Ref.\,\cite{Bernal:2022pue} shows that
\begin{equation}
    \gamma = \frac{3T^6}{4\pi^5 v_{EW}^4} \mathrm{Tr}[m_\nu^\dagger m_\nu]\,.
\end{equation}
When the comoving number density of $B-L$ freezes in when the sphaleron processes go out of equilibrium at $T = T_{\rm sph}$~\cite{DOnofrio:2014rug}. We quantify the final baryon asymmetry as 
\begin{equation}\label{YB}
    |\YB| = \eta_{\rm sph} \frac{|\YBL|}{\mathcal{S}}\,.
\end{equation}
where $\mathcal{S}$ is the comoving entropy density of the Universe obtained by solving Eq.\,\eqref{eq:entropy} and $\eta_{\rm sph}=12/37$ is the sphaleron efficiency factor~\cite{Harvey:1990qw}. 

In Fig.s~\ref{fig:benchmarks_1} and \ref{fig:benchmarks_2}, we present two benchmark solutions for the Boltzmann and Friedmann equations under consideration, in case of the semiclassical Hawking evaporation with $k=0$ (left plots) and the memory-burden effect with $k=1$ and $q=1/2$ (right plots). In both figures, we show the the evolution of thermal and nonthermal abundances of the right-handed neutrinos and the absolute value of the $B-L$ yield (upper panels), the evolution of the PBH mass (middle panels), and the normalized radiation and PBH energy densities (lower panels), as a function of the auxiliary variable $z=M_1/T$. These plots highlight that the memory-burden effect extends the PBHs lifespan, implying a two-phase evaporation. As a result, for certain values of $\beta^\prime$, where they would not have otherwise dominated the Universe, they instead become the dominant component (see e.g. Fig.~\ref{fig:benchmarks_2}).
\begin{figure*}[tbh!]
    \centering
    \includegraphics[width=0.49\linewidth]{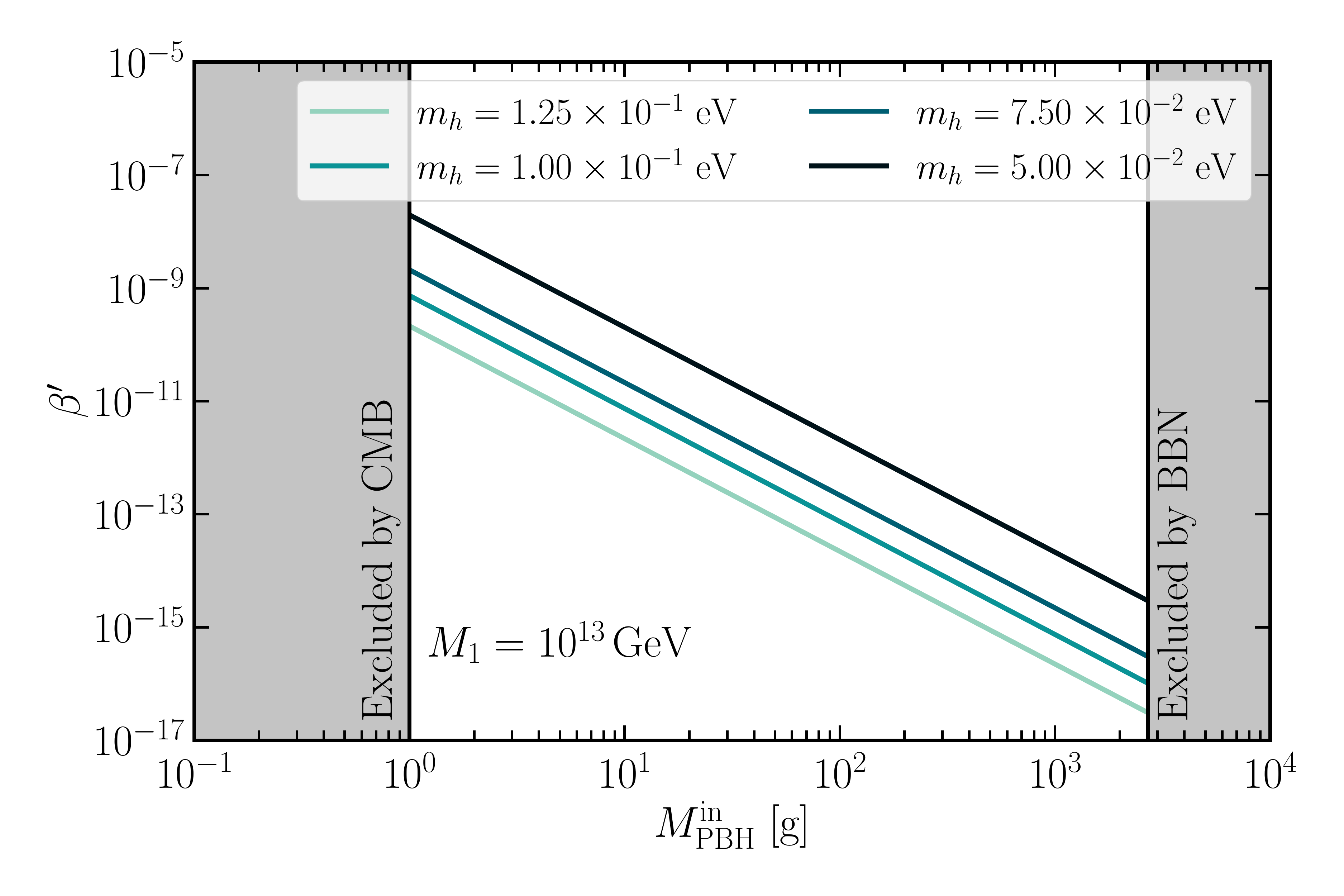}
    \includegraphics[width=0.49\linewidth]{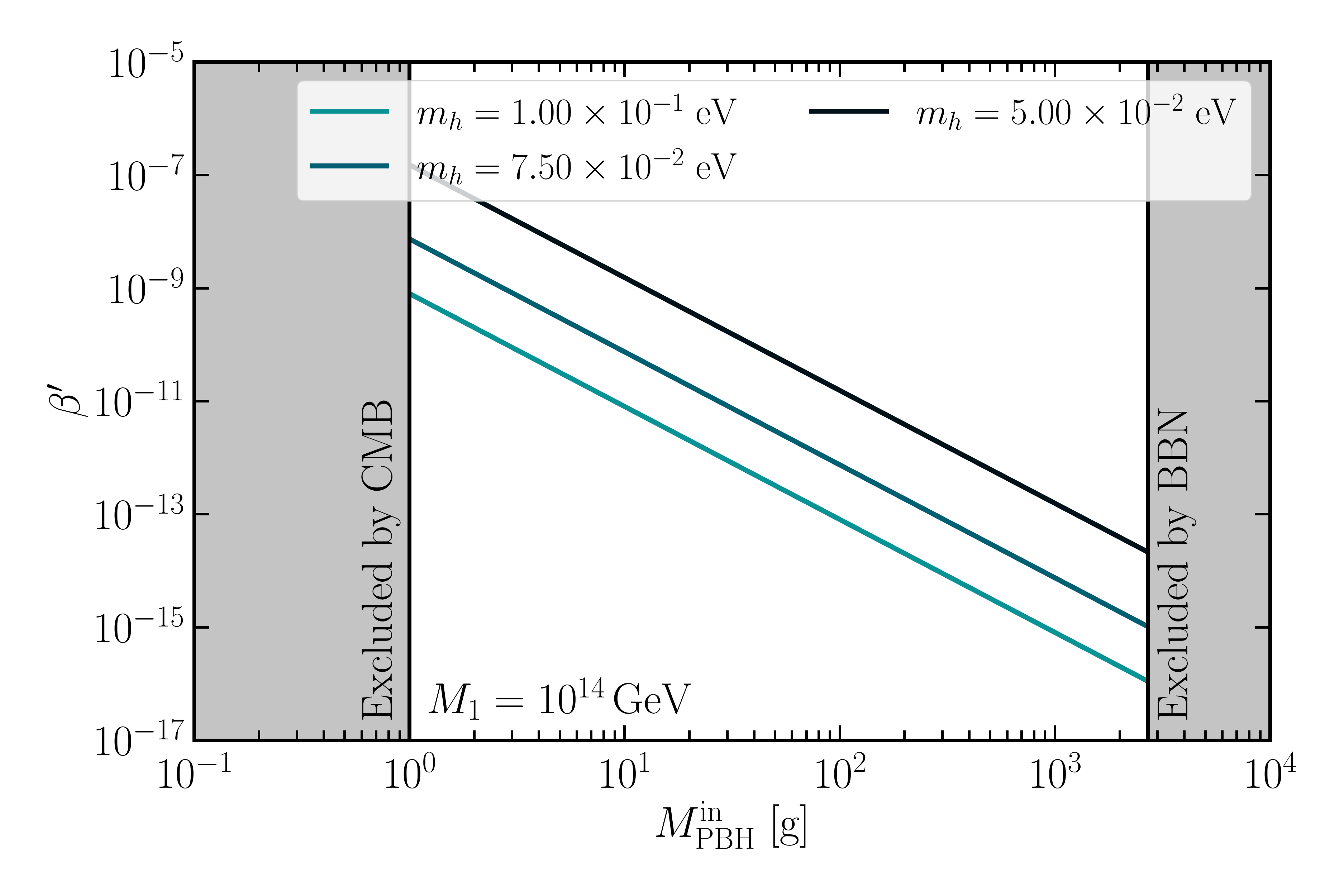}
    \caption{Upper limits on $\beta^{\prime}$ as a function of the initial PBH mass for different values of the heaviest active neutrino mass $m_h$ (shown with different colors). The left (right) plot corresponds to $M_1 = 10^{13}$\,GeV ($M_1 = 10^{14}$\,GeV) of the lightest right-handed neutrinos. The gray bands on the left (right) represent the constraints placed by CMB (BBN) observations according to Ref.~\cite{Haque:2024eyh}.}
    \label{fig:beta_vs_MPBH}
\end{figure*}

As can be seen from these benchmark solutions, the primary effect of PBHs is not their role as a nonthermal source of right-handed neutrinos that would enhance the production of the baryon asymmetry as pointed out by Ref.s~\cite{Bernal:2022pue, Perez-Gonzalez:2020vnz}. Indeed, in the later stages of PBH evaporation, the production of right-handed neutrinos is significantly suppressed and contributes much less to the overall baryon asymmetry. Remarkably, the dominant effect of memory-burdened PBHs stems from the entropy injection, which results in a net reduction of the baryon asymmetry. It is interesting to note that PBHs, with masses in the range $[1,\,1000]$\,g ,exhibit a behavior analogous to that obtained in Ref.~\cite{Calabrese:2023key} for larger masses, in the range $[10^6,\,10^9]$\,g, where the memory-burden effect was not considered.

\section{Results}\label{sec:res}

We here discuss the main results of our analysis obtained by performing a comprehensive scan of the parameter space for $q=1/2$ and $k = 1$ to investigate how the memory-burden effect alters the mixed scenario involving PBHs and high-scale leptogenesis. In Fig.~\ref{fig:beta_vs_MPBH}, we present the upper limits on $\beta^{\prime}$ as a function of the initial PBH mass in case of $M_1 = 10^{13}$\,GeV (left plot) and $M_1 = 10^{14}$\,GeV (right plot). Higher values of $\beta^{\prime}$ are indeed ruled out due to a high entropy injection from memory-burdened evaporation of PBHs which dilutes the baryon asymmetry of the Universe produced by high-scale leptogenesis to values smaller than the observed one~\cite{Planck:2018jri}. Hence, the regions above the lines reported in the figure represent mutual exclusion limits of memory-burdened PBHs and high-scale leptogenesis. The limits are valid for initial PBH masses restricted to the range $[1,\, 1000]~{\rm g}$. Indeed, lighter PBHs are excluded by the constraints on the maximum allowed value of the inflationary energy scale according to CMB measurements, while heavier PBHs are ruled out by Big Bang Nucleosynthesis for $k=1$~\cite{Haque:2024eyh}. Moreover, we find that, as in the standard scenario~\cite{Calabrese:2023key} (see also Ref.s~\cite{Bernal:2022pue, Perez-Gonzalez:2020vnz}), the upper limit on $\beta^{\prime}$ remarkably depends on the heaviest active neutrino mass $m_h$.
\begin{figure*}[tbh!]
    \centering
    \includegraphics[width=0.49\linewidth]{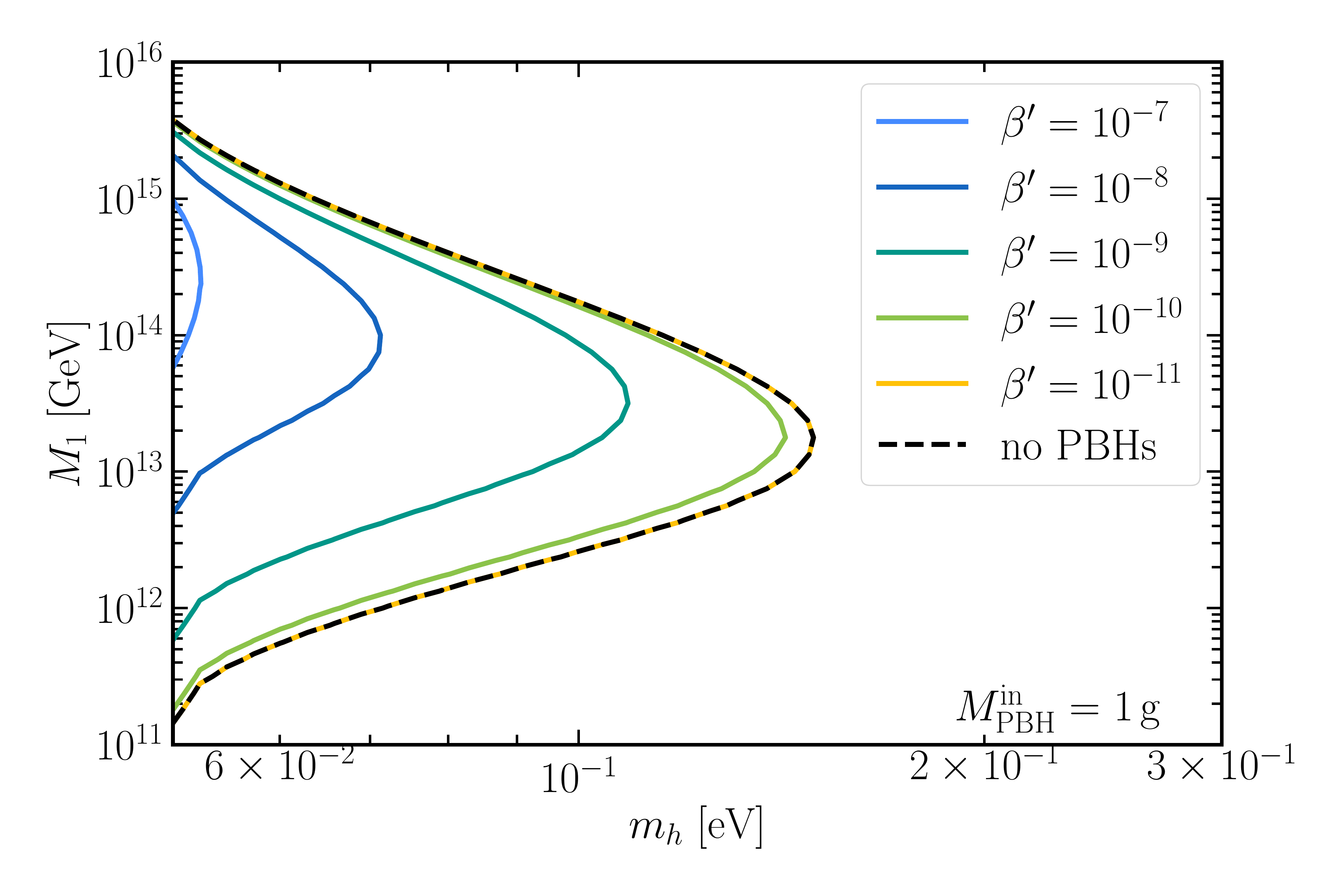}
    \includegraphics[width=0.49\linewidth]{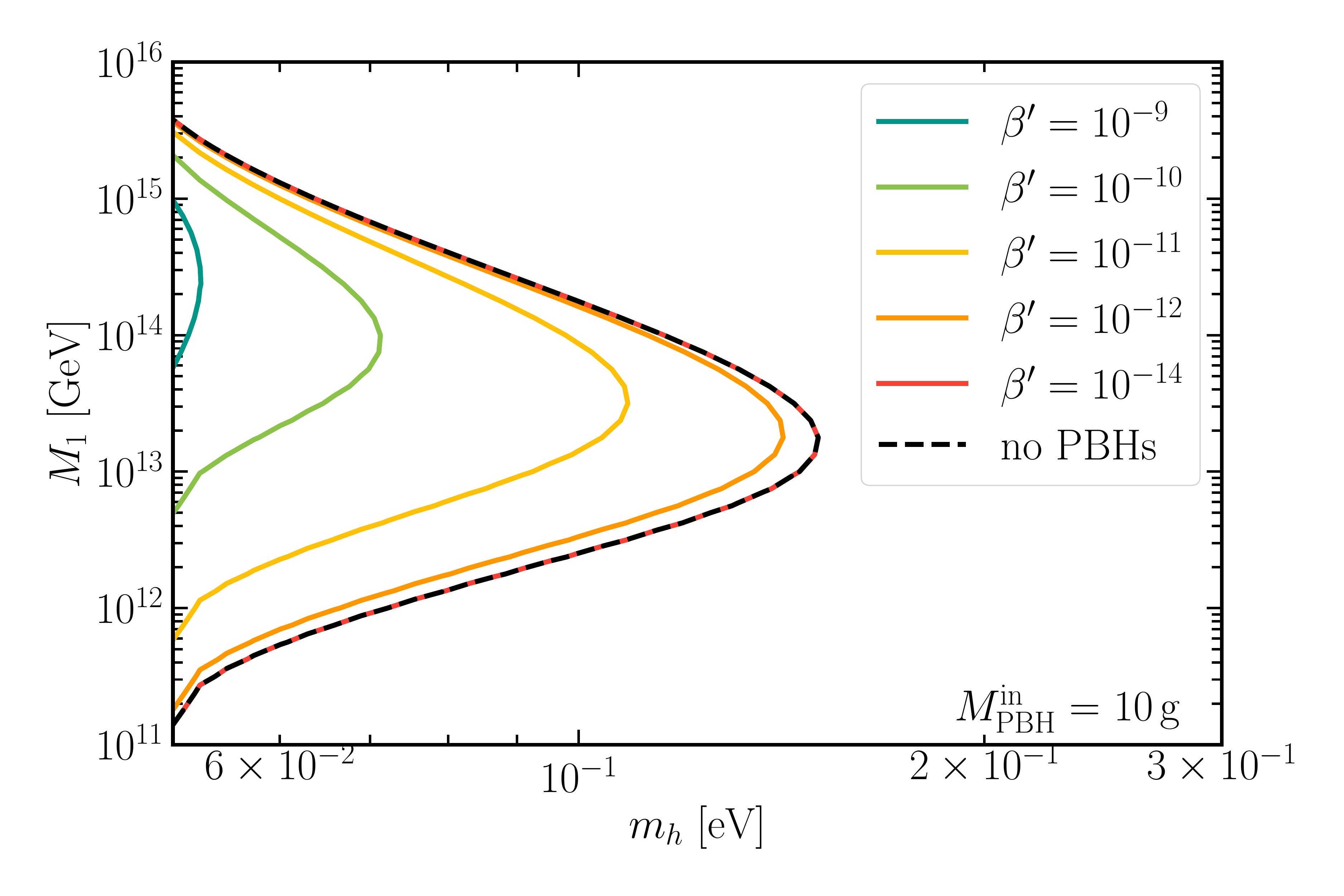}
    \includegraphics[width=0.49\linewidth]{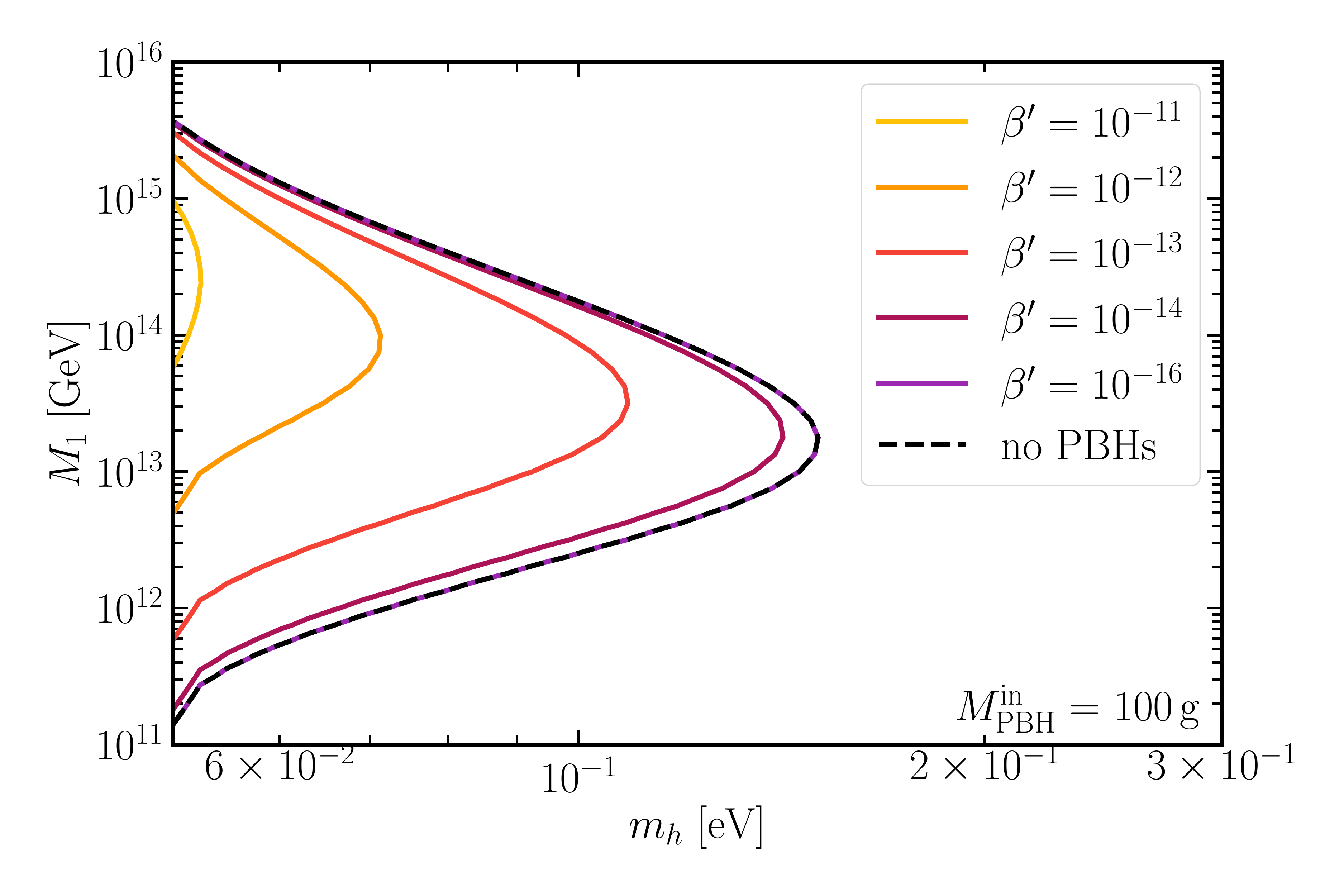}
    \includegraphics[width=0.49\linewidth]{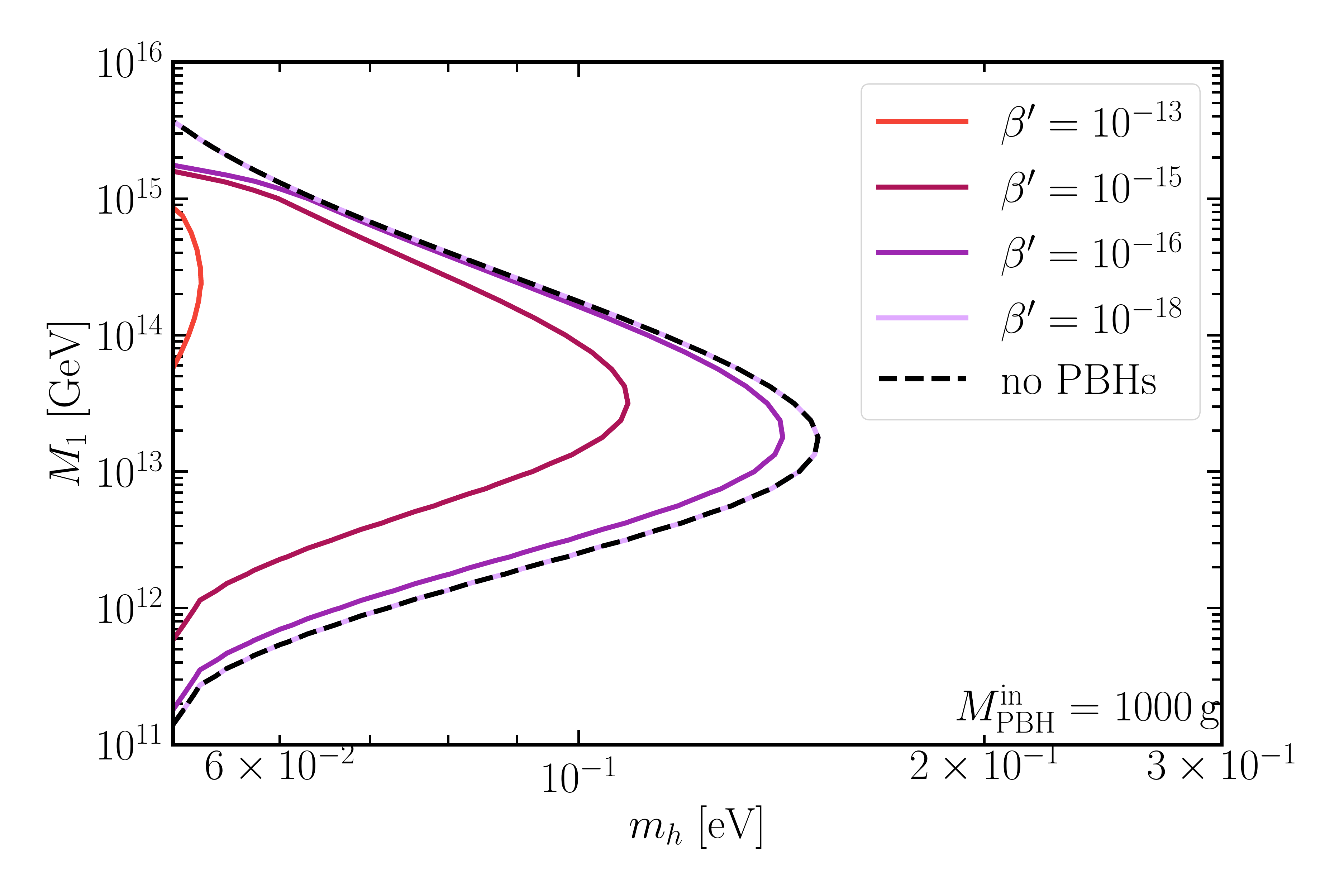}
    \caption{Contour lines in the plane $m_h$--$M_1$ for which the observed baryon asymmetry can be reproduced in case of different values for the PBH abundance $\beta^\prime$ (shown with different colors). The lines delimit to the right the regions where the baryon asymmetry is higher than the observed value. The panels correspond to different assumptions for the initial PBH mass.}
    \label{fig:mh_vs_mN}
\end{figure*}

In Fig.~\ref{fig:mh_vs_mN}, we instead show the regions in the $m_h$--$M_1$ plane where the observed baryon asymmetry can be successfully reproduced for three benchmark values of the initial PBH mass ($\MPBH = 1,\,10,\,100$\,g). The colored lines indicate the parameters achieving the observed baryon asymmetry in case of different assumed values for the PBH abundance $\beta^\prime$. These lines can be compared with the black dashed line corresponding to the standard leptogenesis scenario without memory-burdened PBHs. Higher values of the PBH abundance $\beta^{\prime}$ result in shrinking the viable region to the left due to a higher dilution from entropy injection.

From the results of the present analysis, we observe that a future determination of the absolute neutrino mass from direct measurements in laboratory experiments and from cosmological analyses will provide a strong constraints on memory-burdened PBHs in the framework of high-scale leptogensis.

\section{Conclusions}\label{sec:concl}

We have investigated the impact of a nonstandard cosmology driven by the presence and evaporation of light memory-burdened PBHs on the production of the baryon asymmetry of the Universe through high-scale leptogenesis. We have found that the primary contribution of PBHs is linked to entropy injection, whereas, in the absence of the memory-burden effect, the dominant contribution arises from the additional nonthermal source term of right-handed neutrinos. In this framework, we have explored the mixed parameter space assuming $q=1/2$ and $k = 1$, deriving the mutual exclusion limits and projecting them in the PBHs and high-scale leptogenesis parameter spaces. We have demonstrated that the memory-burden effect significantly reshapes the parameter space, pushing the viable region to lower PBH masses compared to previous studies. This has important implications for understanding baryogenesis in nonstandard cosmologies.

\section*{Acknowledgments}

We thank Stefano Morisi for useful comments and discussions. RC work was supported by the Italian grant 2022JCYC9E ``ReD+, a low-energy characterization for low-mass Dark Matter searches with Argon” (PRIN2022) funded by the Italian Ministero dell’Università e della Ricerca (MUR). MC and NS acknowledge the support by the research project TAsP (Theoretical Astroparticle Physics) funded by the Istituto Nazionale di Fisica Nucleare (INFN).
The work of NS is further supported by the research grant number 2022E2J4RK ``PANTHEON: Perspectives in Astroparticle and Neutrino THEory with Old and New messengers'' under the program PRIN 2022 funded by the Italian Ministero dell’Università e della Ricerca (MUR).

\bibliography{bibliography}

\begin{thebibliography}{84}%
\makeatletter
\providecommand \@ifxundefined [1]{%
 \@ifx{#1\undefined}
}%
\providecommand \@ifnum [1]{%
 \ifnum #1\expandafter \@firstoftwo
 \else \expandafter \@secondoftwo
 \fi
}%
\providecommand \@ifx [1]{%
 \ifx #1\expandafter \@firstoftwo
 \else \expandafter \@secondoftwo
 \fi
}%
\providecommand \natexlab [1]{#1}%
\providecommand \enquote  [1]{``#1''}%
\providecommand \bibnamefont  [1]{#1}%
\providecommand \bibfnamefont [1]{#1}%
\providecommand \citenamefont [1]{#1}%
\providecommand \href@noop [0]{\@secondoftwo}%
\providecommand \href [0]{\begingroup \@sanitize@url \@href}%
\providecommand \@href[1]{\@@startlink{#1}\@@href}%
\providecommand \@@href[1]{\endgroup#1\@@endlink}%
\providecommand \@sanitize@url [0]{\catcode `\\12\catcode `\$12\catcode
  `\&12\catcode `\#12\catcode `\^12\catcode `\_12\catcode `\%12\relax}%
\providecommand \@@startlink[1]{}%
\providecommand \@@endlink[0]{}%
\providecommand \url  [0]{\begingroup\@sanitize@url \@url }%
\providecommand \@url [1]{\endgroup\@href {#1}{\urlprefix }}%
\providecommand \urlprefix  [0]{URL }%
\providecommand \Eprint [0]{\href }%
\providecommand \doibase [0]{http://dx.doi.org/}%
\providecommand \selectlanguage [0]{\@gobble}%
\providecommand \bibinfo  [0]{\@secondoftwo}%
\providecommand \bibfield  [0]{\@secondoftwo}%
\providecommand \translation [1]{[#1]}%
\providecommand \BibitemOpen [0]{}%
\providecommand \bibitemStop [0]{}%
\providecommand \bibitemNoStop [0]{.\EOS\space}%
\providecommand \EOS [0]{\spacefactor3000\relax}%
\providecommand \BibitemShut  [1]{\csname bibitem#1\endcsname}%
\let\auto@bib@innerbib\@empty
\bibitem [{\citenamefont {Zel'dovich}\ and\ \citenamefont
  {Novikov}(1967)}]{Zeldovich:1967lct}%
  \BibitemOpen
  \bibfield  {author} {\bibinfo {author} {\bibfnamefont {Ya.~B.}\ \bibnamefont
  {Zel'dovich}}\ and\ \bibinfo {author} {\bibfnamefont {I.~D.}\ \bibnamefont
  {Novikov}},\ }\bibfield  {title} {\enquote {\bibinfo {title} {{The Hypothesis
  of Cores Retarded during Expansion and the Hot Cosmological Model}},}\
  }\href@noop {} {\bibfield  {journal} {\bibinfo  {journal} {Soviet Astron. AJ
  (Engl. Transl. ),}\ }\textbf {\bibinfo {volume} {10}},\ \bibinfo {pages}
  {602} (\bibinfo {year} {1967})}\BibitemShut {NoStop}%
\bibitem [{\citenamefont {Hawking}(1971)}]{Hawking:1971ei}%
  \BibitemOpen
  \bibfield  {author} {\bibinfo {author} {\bibfnamefont {Stephen}\ \bibnamefont
  {Hawking}},\ }\bibfield  {title} {\enquote {\bibinfo {title}
  {{Gravitationally collapsed objects of very low mass}},}\ }\href@noop {}
  {\bibfield  {journal} {\bibinfo  {journal} {Mon. Not. Roy. Astron. Soc.}\
  }\textbf {\bibinfo {volume} {152}},\ \bibinfo {pages} {75} (\bibinfo {year}
  {1971})}\BibitemShut {NoStop}%
\bibitem [{\citenamefont {Carr}\ and\ \citenamefont
  {Hawking}(1974)}]{Carr:1974nx}%
  \BibitemOpen
  \bibfield  {author} {\bibinfo {author} {\bibfnamefont {Bernard~J.}\
  \bibnamefont {Carr}}\ and\ \bibinfo {author} {\bibfnamefont {S.~W.}\
  \bibnamefont {Hawking}},\ }\bibfield  {title} {\enquote {\bibinfo {title}
  {{Black holes in the early Universe}},}\ }\href {\doibase
  10.1093/mnras/168.2.399} {\bibfield  {journal} {\bibinfo  {journal} {Mon.
  Not. Roy. Astron. Soc.}\ }\textbf {\bibinfo {volume} {168}},\ \bibinfo
  {pages} {399--415} (\bibinfo {year} {1974})}\BibitemShut {NoStop}%
\bibitem [{\citenamefont {Bernal}\ and\ \citenamefont
  {Zapata}(2021{\natexlab{a}})}]{Bernal:2020kse}%
  \BibitemOpen
  \bibfield  {author} {\bibinfo {author} {\bibfnamefont {Nicol\'as}\
  \bibnamefont {Bernal}}\ and\ \bibinfo {author} {\bibfnamefont {\'Oscar}\
  \bibnamefont {Zapata}},\ }\bibfield  {title} {\enquote {\bibinfo {title}
  {{Self-interacting Dark Matter from Primordial Black Holes}},}\ }\href
  {\doibase 10.1088/1475-7516/2021/03/007} {\bibfield  {journal} {\bibinfo
  {journal} {JCAP}\ }\textbf {\bibinfo {volume} {03}},\ \bibinfo {pages} {007}
  (\bibinfo {year} {2021}{\natexlab{a}})},\ \Eprint
  {http://arxiv.org/abs/2010.09725} {arXiv:2010.09725 [hep-ph]} \BibitemShut
  {NoStop}%
\bibitem [{\citenamefont {Gondolo}\ \emph {et~al.}(2020)\citenamefont
  {Gondolo}, \citenamefont {Sandick},\ and\ \citenamefont {Shams
  Es~Haghi}}]{Gondolo:2020uqv}%
  \BibitemOpen
  \bibfield  {author} {\bibinfo {author} {\bibfnamefont {Paolo}\ \bibnamefont
  {Gondolo}}, \bibinfo {author} {\bibfnamefont {Pearl}\ \bibnamefont
  {Sandick}}, \ and\ \bibinfo {author} {\bibfnamefont {Barmak}\ \bibnamefont
  {Shams Es~Haghi}},\ }\bibfield  {title} {\enquote {\bibinfo {title} {{Effects
  of primordial black holes on dark matter models}},}\ }\href {\doibase
  10.1103/PhysRevD.102.095018} {\bibfield  {journal} {\bibinfo  {journal}
  {Phys. Rev. D}\ }\textbf {\bibinfo {volume} {102}},\ \bibinfo {pages}
  {095018} (\bibinfo {year} {2020})},\ \Eprint
  {http://arxiv.org/abs/2009.02424} {arXiv:2009.02424 [hep-ph]} \BibitemShut
  {NoStop}%
\bibitem [{\citenamefont {Bernal}\ and\ \citenamefont
  {Zapata}(2021{\natexlab{b}})}]{Bernal:2020ili}%
  \BibitemOpen
  \bibfield  {author} {\bibinfo {author} {\bibfnamefont {Nicol\'as}\
  \bibnamefont {Bernal}}\ and\ \bibinfo {author} {\bibfnamefont {\'Oscar}\
  \bibnamefont {Zapata}},\ }\bibfield  {title} {\enquote {\bibinfo {title}
  {{Gravitational dark matter production: primordial black holes and UV
  freeze-in}},}\ }\href {\doibase 10.1016/j.physletb.2021.136129} {\bibfield
  {journal} {\bibinfo  {journal} {Phys. Lett. B}\ }\textbf {\bibinfo {volume}
  {815}},\ \bibinfo {pages} {136129} (\bibinfo {year} {2021}{\natexlab{b}})},\
  \Eprint {http://arxiv.org/abs/2011.02510} {arXiv:2011.02510 [hep-ph]}
  \BibitemShut {NoStop}%
\bibitem [{\citenamefont {Bernal}\ and\ \citenamefont
  {Zapata}(2021{\natexlab{c}})}]{Bernal:2020bjf}%
  \BibitemOpen
  \bibfield  {author} {\bibinfo {author} {\bibfnamefont {Nicol\'as}\
  \bibnamefont {Bernal}}\ and\ \bibinfo {author} {\bibfnamefont {\'Oscar}\
  \bibnamefont {Zapata}},\ }\bibfield  {title} {\enquote {\bibinfo {title}
  {{Dark Matter in the Time of Primordial Black Holes}},}\ }\href {\doibase
  10.1088/1475-7516/2021/03/015} {\bibfield  {journal} {\bibinfo  {journal}
  {JCAP}\ }\textbf {\bibinfo {volume} {03}},\ \bibinfo {pages} {015} (\bibinfo
  {year} {2021}{\natexlab{c}})},\ \Eprint {http://arxiv.org/abs/2011.12306}
  {arXiv:2011.12306 [astro-ph.CO]} \BibitemShut {NoStop}%
\bibitem [{\citenamefont {Cheek}\ \emph
  {et~al.}(2022{\natexlab{a}})\citenamefont {Cheek}, \citenamefont {Heurtier},
  \citenamefont {Perez-Gonzalez},\ and\ \citenamefont
  {Turner}}]{Cheek:2021odj}%
  \BibitemOpen
  \bibfield  {author} {\bibinfo {author} {\bibfnamefont {Andrew}\ \bibnamefont
  {Cheek}}, \bibinfo {author} {\bibfnamefont {Lucien}\ \bibnamefont
  {Heurtier}}, \bibinfo {author} {\bibfnamefont {Yuber~F.}\ \bibnamefont
  {Perez-Gonzalez}}, \ and\ \bibinfo {author} {\bibfnamefont {Jessica}\
  \bibnamefont {Turner}},\ }\bibfield  {title} {\enquote {\bibinfo {title}
  {{Primordial black hole evaporation and dark matter production. I. Solely
  Hawking radiation}},}\ }\href {\doibase 10.1103/PhysRevD.105.015022}
  {\bibfield  {journal} {\bibinfo  {journal} {Phys. Rev. D}\ }\textbf {\bibinfo
  {volume} {105}},\ \bibinfo {pages} {015022} (\bibinfo {year}
  {2022}{\natexlab{a}})},\ \Eprint {http://arxiv.org/abs/2107.00013}
  {arXiv:2107.00013 [hep-ph]} \BibitemShut {NoStop}%
\bibitem [{\citenamefont {Cheek}\ \emph
  {et~al.}(2022{\natexlab{b}})\citenamefont {Cheek}, \citenamefont {Heurtier},
  \citenamefont {Perez-Gonzalez},\ and\ \citenamefont
  {Turner}}]{Cheek:2021cfe}%
  \BibitemOpen
  \bibfield  {author} {\bibinfo {author} {\bibfnamefont {Andrew}\ \bibnamefont
  {Cheek}}, \bibinfo {author} {\bibfnamefont {Lucien}\ \bibnamefont
  {Heurtier}}, \bibinfo {author} {\bibfnamefont {Yuber~F.}\ \bibnamefont
  {Perez-Gonzalez}}, \ and\ \bibinfo {author} {\bibfnamefont {Jessica}\
  \bibnamefont {Turner}},\ }\bibfield  {title} {\enquote {\bibinfo {title}
  {{Primordial black hole evaporation and dark matter production. II. Interplay
  with the freeze-in or freeze-out mechanism}},}\ }\href {\doibase
  10.1103/PhysRevD.105.015023} {\bibfield  {journal} {\bibinfo  {journal}
  {Phys. Rev. D}\ }\textbf {\bibinfo {volume} {105}},\ \bibinfo {pages}
  {015023} (\bibinfo {year} {2022}{\natexlab{b}})},\ \Eprint
  {http://arxiv.org/abs/2107.00016} {arXiv:2107.00016 [hep-ph]} \BibitemShut
  {NoStop}%
\bibitem [{\citenamefont {Samanta}\ and\ \citenamefont
  {Urban}(2022)}]{Samanta:2021mdm}%
  \BibitemOpen
  \bibfield  {author} {\bibinfo {author} {\bibfnamefont {Rome}\ \bibnamefont
  {Samanta}}\ and\ \bibinfo {author} {\bibfnamefont {Federico~R.}\ \bibnamefont
  {Urban}},\ }\bibfield  {title} {\enquote {\bibinfo {title} {{Testing super
  heavy dark matter from primordial black holes with gravitational waves}},}\
  }\href {\doibase 10.1088/1475-7516/2022/06/017} {\bibfield  {journal}
  {\bibinfo  {journal} {JCAP}\ }\textbf {\bibinfo {volume} {06}},\ \bibinfo
  {pages} {017} (\bibinfo {year} {2022})},\ \Eprint
  {http://arxiv.org/abs/2112.04836} {arXiv:2112.04836 [hep-ph]} \BibitemShut
  {NoStop}%
\bibitem [{\citenamefont {Bernal}\ \emph
  {et~al.}(2021{\natexlab{a}})\citenamefont {Bernal}, \citenamefont
  {Hajkarim},\ and\ \citenamefont {Xu}}]{Bernal:2021yyb}%
  \BibitemOpen
  \bibfield  {author} {\bibinfo {author} {\bibfnamefont {Nicol\'as}\
  \bibnamefont {Bernal}}, \bibinfo {author} {\bibfnamefont {Fazlollah}\
  \bibnamefont {Hajkarim}}, \ and\ \bibinfo {author} {\bibfnamefont {Yong}\
  \bibnamefont {Xu}},\ }\bibfield  {title} {\enquote {\bibinfo {title} {{Axion
  Dark Matter in the Time of Primordial Black Holes}},}\ }\href {\doibase
  10.1103/PhysRevD.104.075007} {\bibfield  {journal} {\bibinfo  {journal}
  {Phys. Rev. D}\ }\textbf {\bibinfo {volume} {104}},\ \bibinfo {pages}
  {075007} (\bibinfo {year} {2021}{\natexlab{a}})},\ \Eprint
  {http://arxiv.org/abs/2107.13575} {arXiv:2107.13575 [hep-ph]} \BibitemShut
  {NoStop}%
\bibitem [{\citenamefont {Bernal}\ \emph
  {et~al.}(2021{\natexlab{b}})\citenamefont {Bernal}, \citenamefont
  {Perez-Gonzalez}, \citenamefont {Xu},\ and\ \citenamefont
  {Zapata}}]{Bernal:2021bbv}%
  \BibitemOpen
  \bibfield  {author} {\bibinfo {author} {\bibfnamefont {Nicol\'as}\
  \bibnamefont {Bernal}}, \bibinfo {author} {\bibfnamefont {Yuber~F.}\
  \bibnamefont {Perez-Gonzalez}}, \bibinfo {author} {\bibfnamefont {Yong}\
  \bibnamefont {Xu}}, \ and\ \bibinfo {author} {\bibfnamefont {\'Oscar}\
  \bibnamefont {Zapata}},\ }\bibfield  {title} {\enquote {\bibinfo {title}
  {{ALP dark matter in a primordial black hole dominated universe}},}\ }\href
  {\doibase 10.1103/PhysRevD.104.123536} {\bibfield  {journal} {\bibinfo
  {journal} {Phys. Rev. D}\ }\textbf {\bibinfo {volume} {104}},\ \bibinfo
  {pages} {123536} (\bibinfo {year} {2021}{\natexlab{b}})},\ \Eprint
  {http://arxiv.org/abs/2110.04312} {arXiv:2110.04312 [hep-ph]} \BibitemShut
  {NoStop}%
\bibitem [{\citenamefont {Sandick}\ \emph {et~al.}(2021)\citenamefont
  {Sandick}, \citenamefont {Es~Haghi},\ and\ \citenamefont
  {Sinha}}]{Sandick:2021gew}%
  \BibitemOpen
  \bibfield  {author} {\bibinfo {author} {\bibfnamefont {Pearl}\ \bibnamefont
  {Sandick}}, \bibinfo {author} {\bibfnamefont {Barmak~Shams}\ \bibnamefont
  {Es~Haghi}}, \ and\ \bibinfo {author} {\bibfnamefont {Kuver}\ \bibnamefont
  {Sinha}},\ }\bibfield  {title} {\enquote {\bibinfo {title} {{Asymmetric
  reheating by primordial black holes}},}\ }\href {\doibase
  10.1103/PhysRevD.104.083523} {\bibfield  {journal} {\bibinfo  {journal}
  {Phys. Rev. D}\ }\textbf {\bibinfo {volume} {104}},\ \bibinfo {pages}
  {083523} (\bibinfo {year} {2021})},\ \Eprint
  {http://arxiv.org/abs/2108.08329} {arXiv:2108.08329 [astro-ph.CO]}
  \BibitemShut {NoStop}%
\bibitem [{\citenamefont {Bernal}\ \emph
  {et~al.}(2022{\natexlab{a}})\citenamefont {Bernal}, \citenamefont
  {Perez-Gonzalez},\ and\ \citenamefont {Xu}}]{Bernal:2022oha}%
  \BibitemOpen
  \bibfield  {author} {\bibinfo {author} {\bibfnamefont {Nicol\'as}\
  \bibnamefont {Bernal}}, \bibinfo {author} {\bibfnamefont {Yuber~F.}\
  \bibnamefont {Perez-Gonzalez}}, \ and\ \bibinfo {author} {\bibfnamefont
  {Yong}\ \bibnamefont {Xu}},\ }\bibfield  {title} {\enquote {\bibinfo {title}
  {{Superradiant production of heavy dark matter from primordial black
  holes}},}\ }\href {\doibase 10.1103/PhysRevD.106.015020} {\bibfield
  {journal} {\bibinfo  {journal} {Phys. Rev. D}\ }\textbf {\bibinfo {volume}
  {106}},\ \bibinfo {pages} {015020} (\bibinfo {year} {2022}{\natexlab{a}})},\
  \Eprint {http://arxiv.org/abs/2205.11522} {arXiv:2205.11522 [hep-ph]}
  \BibitemShut {NoStop}%
\bibitem [{\citenamefont {Cheek}\ \emph {et~al.}(2023)\citenamefont {Cheek},
  \citenamefont {Heurtier}, \citenamefont {Perez-Gonzalez},\ and\ \citenamefont
  {Turner}}]{Cheek:2022mmy}%
  \BibitemOpen
  \bibfield  {author} {\bibinfo {author} {\bibfnamefont {Andrew}\ \bibnamefont
  {Cheek}}, \bibinfo {author} {\bibfnamefont {Lucien}\ \bibnamefont
  {Heurtier}}, \bibinfo {author} {\bibfnamefont {Yuber~F.}\ \bibnamefont
  {Perez-Gonzalez}}, \ and\ \bibinfo {author} {\bibfnamefont {Jessica}\
  \bibnamefont {Turner}},\ }\bibfield  {title} {\enquote {\bibinfo {title}
  {{Evaporation of primordial black holes in the early Universe: Mass and spin
  distributions}},}\ }\href {\doibase 10.1103/PhysRevD.108.015005} {\bibfield
  {journal} {\bibinfo  {journal} {Phys. Rev. D}\ }\textbf {\bibinfo {volume}
  {108}},\ \bibinfo {pages} {015005} (\bibinfo {year} {2023})},\ \Eprint
  {http://arxiv.org/abs/2212.03878} {arXiv:2212.03878 [hep-ph]} \BibitemShut
  {NoStop}%
\bibitem [{\citenamefont {Gehrman}\ \emph {et~al.}(2024)\citenamefont
  {Gehrman}, \citenamefont {Shams Es~Haghi}, \citenamefont {Sinha},\ and\
  \citenamefont {Xu}}]{Gehrman:2023qjn}%
  \BibitemOpen
  \bibfield  {author} {\bibinfo {author} {\bibfnamefont {Thomas~C.}\
  \bibnamefont {Gehrman}}, \bibinfo {author} {\bibfnamefont {Barmak}\
  \bibnamefont {Shams Es~Haghi}}, \bibinfo {author} {\bibfnamefont {Kuver}\
  \bibnamefont {Sinha}}, \ and\ \bibinfo {author} {\bibfnamefont {Tao}\
  \bibnamefont {Xu}},\ }\bibfield  {title} {\enquote {\bibinfo {title}
  {{Recycled dark matter}},}\ }\href {\doibase 10.1088/1475-7516/2024/03/044}
  {\bibfield  {journal} {\bibinfo  {journal} {JCAP}\ }\textbf {\bibinfo
  {volume} {03}},\ \bibinfo {pages} {044} (\bibinfo {year} {2024})},\ \Eprint
  {http://arxiv.org/abs/2310.08526} {arXiv:2310.08526 [hep-ph]} \BibitemShut
  {NoStop}%
\bibitem [{\citenamefont {Bertuzzo}\ \emph {et~al.}(2024)\citenamefont
  {Bertuzzo}, \citenamefont {Perez-Gonzalez}, \citenamefont {Salla},\ and\
  \citenamefont {Funchal}}]{Bertuzzo:2024fns}%
  \BibitemOpen
  \bibfield  {author} {\bibinfo {author} {\bibfnamefont {Enrico}\ \bibnamefont
  {Bertuzzo}}, \bibinfo {author} {\bibfnamefont {Yuber~F.}\ \bibnamefont
  {Perez-Gonzalez}}, \bibinfo {author} {\bibfnamefont {Gabriel~M.}\
  \bibnamefont {Salla}}, \ and\ \bibinfo {author} {\bibfnamefont
  {Renata~Zukanovich}\ \bibnamefont {Funchal}},\ }\bibfield  {title} {\enquote
  {\bibinfo {title} {{Gravitationally produced dark matter and primordial black
  holes}},}\ }\href {\doibase 10.1088/1475-7516/2024/09/059} {\bibfield
  {journal} {\bibinfo  {journal} {JCAP}\ }\textbf {\bibinfo {volume} {09}},\
  \bibinfo {pages} {059} (\bibinfo {year} {2024})},\ \Eprint
  {http://arxiv.org/abs/2405.17611} {arXiv:2405.17611 [hep-ph]} \BibitemShut
  {NoStop}%
\bibitem [{\citenamefont {Basumatary}\ \emph {et~al.}(2024)\citenamefont
  {Basumatary}, \citenamefont {Raj},\ and\ \citenamefont
  {Ray}}]{Basumatary:2024uwo}%
  \BibitemOpen
  \bibfield  {author} {\bibinfo {author} {\bibfnamefont {Ujjwal}\ \bibnamefont
  {Basumatary}}, \bibinfo {author} {\bibfnamefont {Nirmal}\ \bibnamefont
  {Raj}}, \ and\ \bibinfo {author} {\bibfnamefont {Anupam}\ \bibnamefont
  {Ray}},\ }\bibfield  {title} {\enquote {\bibinfo {title} {{Beyond Hawking
  evaporation of black holes formed by dark matter in compact stars}},}\
  }\href@noop {} {\  (\bibinfo {year} {2024})},\ \Eprint
  {http://arxiv.org/abs/2410.22702} {arXiv:2410.22702 [hep-ph]} \BibitemShut
  {NoStop}%
\bibitem [{\citenamefont {Dom\`enech}\ and\ \citenamefont
  {Tr\"ankle}(2024)}]{Domenech:2024wao}%
  \BibitemOpen
  \bibfield  {author} {\bibinfo {author} {\bibfnamefont {Guillem}\ \bibnamefont
  {Dom\`enech}}\ and\ \bibinfo {author} {\bibfnamefont {Jan}\ \bibnamefont
  {Tr\"ankle}},\ }\bibfield  {title} {\enquote {\bibinfo {title} {{From
  formation to evaporation: Induced gravitational wave probes of the primordial
  black hole reheating scenario}},}\ }\href@noop {} {\  (\bibinfo {year}
  {2024})},\ \Eprint {http://arxiv.org/abs/2409.12125} {arXiv:2409.12125
  [gr-qc]} \BibitemShut {NoStop}%
\bibitem [{\citenamefont {Carr}\ \emph {et~al.}(2016)\citenamefont {Carr},
  \citenamefont {Kuhnel},\ and\ \citenamefont {Sandstad}}]{Carr:2016drx}%
  \BibitemOpen
  \bibfield  {author} {\bibinfo {author} {\bibfnamefont {Bernard}\ \bibnamefont
  {Carr}}, \bibinfo {author} {\bibfnamefont {Florian}\ \bibnamefont {Kuhnel}},
  \ and\ \bibinfo {author} {\bibfnamefont {Marit}\ \bibnamefont {Sandstad}},\
  }\bibfield  {title} {\enquote {\bibinfo {title} {{Primordial Black Holes as
  Dark Matter}},}\ }\href {\doibase 10.1103/PhysRevD.94.083504} {\bibfield
  {journal} {\bibinfo  {journal} {Phys. Rev. D}\ }\textbf {\bibinfo {volume}
  {94}},\ \bibinfo {pages} {083504} (\bibinfo {year} {2016})},\ \Eprint
  {http://arxiv.org/abs/1607.06077} {arXiv:1607.06077 [astro-ph.CO]}
  \BibitemShut {NoStop}%
\bibitem [{\citenamefont {Green}\ and\ \citenamefont
  {Kavanagh}(2021)}]{Green:2020jor}%
  \BibitemOpen
  \bibfield  {author} {\bibinfo {author} {\bibfnamefont {Anne~M.}\ \bibnamefont
  {Green}}\ and\ \bibinfo {author} {\bibfnamefont {Bradley~J.}\ \bibnamefont
  {Kavanagh}},\ }\bibfield  {title} {\enquote {\bibinfo {title} {{Primordial
  Black Holes as a dark matter candidate}},}\ }\href {\doibase
  10.1088/1361-6471/abc534} {\bibfield  {journal} {\bibinfo  {journal} {J.
  Phys. G}\ }\textbf {\bibinfo {volume} {48}},\ \bibinfo {pages} {043001}
  (\bibinfo {year} {2021})},\ \Eprint {http://arxiv.org/abs/2007.10722}
  {arXiv:2007.10722 [astro-ph.CO]} \BibitemShut {NoStop}%
\bibitem [{\citenamefont {Carr}\ \emph {et~al.}(2021)\citenamefont {Carr},
  \citenamefont {Kohri}, \citenamefont {Sendouda},\ and\ \citenamefont
  {Yokoyama}}]{Carr:2020gox}%
  \BibitemOpen
  \bibfield  {author} {\bibinfo {author} {\bibfnamefont {Bernard}\ \bibnamefont
  {Carr}}, \bibinfo {author} {\bibfnamefont {Kazunori}\ \bibnamefont {Kohri}},
  \bibinfo {author} {\bibfnamefont {Yuuiti}\ \bibnamefont {Sendouda}}, \ and\
  \bibinfo {author} {\bibfnamefont {Jun'ichi}\ \bibnamefont {Yokoyama}},\
  }\bibfield  {title} {\enquote {\bibinfo {title} {{Constraints on primordial
  black holes}},}\ }\href {\doibase 10.1088/1361-6633/ac1e31} {\bibfield
  {journal} {\bibinfo  {journal} {Rept. Prog. Phys.}\ }\textbf {\bibinfo
  {volume} {84}},\ \bibinfo {pages} {116902} (\bibinfo {year} {2021})},\
  \Eprint {http://arxiv.org/abs/2002.12778} {arXiv:2002.12778 [astro-ph.CO]}
  \BibitemShut {NoStop}%
\bibitem [{\citenamefont {Carr}\ and\ \citenamefont
  {Kuhnel}(2022)}]{Carr:2021bzv}%
  \BibitemOpen
  \bibfield  {author} {\bibinfo {author} {\bibfnamefont {Bernard}\ \bibnamefont
  {Carr}}\ and\ \bibinfo {author} {\bibfnamefont {Florian}\ \bibnamefont
  {Kuhnel}},\ }\bibfield  {title} {\enquote {\bibinfo {title} {{Primordial
  black holes as dark matter candidates}},}\ }\href {\doibase
  10.21468/SciPostPhysLectNotes.48} {\bibfield  {journal} {\bibinfo  {journal}
  {SciPost Phys. Lect. Notes}\ }\textbf {\bibinfo {volume} {48}},\ \bibinfo
  {pages} {1} (\bibinfo {year} {2022})},\ \Eprint
  {http://arxiv.org/abs/2110.02821} {arXiv:2110.02821 [astro-ph.CO]}
  \BibitemShut {NoStop}%
\bibitem [{\citenamefont {Papanikolaou}\ \emph {et~al.}(2021)\citenamefont
  {Papanikolaou}, \citenamefont {Vennin},\ and\ \citenamefont
  {Langlois}}]{Papanikolaou:2020qtd}%
  \BibitemOpen
  \bibfield  {author} {\bibinfo {author} {\bibfnamefont {Theodoros}\
  \bibnamefont {Papanikolaou}}, \bibinfo {author} {\bibfnamefont {Vincent}\
  \bibnamefont {Vennin}}, \ and\ \bibinfo {author} {\bibfnamefont {David}\
  \bibnamefont {Langlois}},\ }\bibfield  {title} {\enquote {\bibinfo {title}
  {{Gravitational waves from a universe filled with primordial black holes}},}\
  }\href {\doibase 10.1088/1475-7516/2021/03/053} {\bibfield  {journal}
  {\bibinfo  {journal} {JCAP}\ }\textbf {\bibinfo {volume} {03}},\ \bibinfo
  {pages} {053} (\bibinfo {year} {2021})},\ \Eprint
  {http://arxiv.org/abs/2010.11573} {arXiv:2010.11573 [astro-ph.CO]}
  \BibitemShut {NoStop}%
\bibitem [{\citenamefont {Dom\`enech}\ \emph {et~al.}(2021)\citenamefont
  {Dom\`enech}, \citenamefont {Lin},\ and\ \citenamefont
  {Sasaki}}]{Domenech:2020ssp}%
  \BibitemOpen
  \bibfield  {author} {\bibinfo {author} {\bibfnamefont {Guillem}\ \bibnamefont
  {Dom\`enech}}, \bibinfo {author} {\bibfnamefont {Chunshan}\ \bibnamefont
  {Lin}}, \ and\ \bibinfo {author} {\bibfnamefont {Misao}\ \bibnamefont
  {Sasaki}},\ }\bibfield  {title} {\enquote {\bibinfo {title} {{Gravitational
  wave constraints on the primordial black hole dominated early universe}},}\
  }\href {\doibase 10.1088/1475-7516/2021/11/E01} {\bibfield  {journal}
  {\bibinfo  {journal} {JCAP}\ }\textbf {\bibinfo {volume} {04}},\ \bibinfo
  {pages} {062} (\bibinfo {year} {2021})},\ \bibinfo {note} {[Erratum: JCAP 11,
  E01 (2021)]},\ \Eprint {http://arxiv.org/abs/2012.08151} {arXiv:2012.08151
  [gr-qc]} \BibitemShut {NoStop}%
\bibitem [{\citenamefont {Papanikolaou}(2022)}]{Papanikolaou:2022chm}%
  \BibitemOpen
  \bibfield  {author} {\bibinfo {author} {\bibfnamefont {Theodoros}\
  \bibnamefont {Papanikolaou}},\ }\bibfield  {title} {\enquote {\bibinfo
  {title} {{Gravitational waves induced from primordial black hole
  fluctuations: the~effect of an extended mass function}},}\ }\href {\doibase
  10.1088/1475-7516/2022/10/089} {\bibfield  {journal} {\bibinfo  {journal}
  {JCAP}\ }\textbf {\bibinfo {volume} {10}},\ \bibinfo {pages} {089} (\bibinfo
  {year} {2022})},\ \Eprint {http://arxiv.org/abs/2207.11041} {arXiv:2207.11041
  [astro-ph.CO]} \BibitemShut {NoStop}%
\bibitem [{\citenamefont {Ireland}\ \emph {et~al.}(2023)\citenamefont
  {Ireland}, \citenamefont {Profumo},\ and\ \citenamefont
  {Scharnhorst}}]{Ireland:2023avg}%
  \BibitemOpen
  \bibfield  {author} {\bibinfo {author} {\bibfnamefont {Aurora}\ \bibnamefont
  {Ireland}}, \bibinfo {author} {\bibfnamefont {Stefano}\ \bibnamefont
  {Profumo}}, \ and\ \bibinfo {author} {\bibfnamefont {Jordan}\ \bibnamefont
  {Scharnhorst}},\ }\bibfield  {title} {\enquote {\bibinfo {title} {{Primordial
  gravitational waves from black hole evaporation in standard and nonstandard
  cosmologies}},}\ }\href {\doibase 10.1103/PhysRevD.107.104021} {\bibfield
  {journal} {\bibinfo  {journal} {Phys. Rev. D}\ }\textbf {\bibinfo {volume}
  {107}},\ \bibinfo {pages} {104021} (\bibinfo {year} {2023})},\ \Eprint
  {http://arxiv.org/abs/2302.10188} {arXiv:2302.10188 [gr-qc]} \BibitemShut
  {NoStop}%
\bibitem [{\citenamefont {Hawking}(1975)}]{Hawking:1975vcx}%
  \BibitemOpen
  \bibfield  {author} {\bibinfo {author} {\bibfnamefont {S.~W.}\ \bibnamefont
  {Hawking}},\ }\bibfield  {title} {\enquote {\bibinfo {title} {{Particle
  Creation by Black Holes}},}\ }\href {\doibase 10.1007/BF02345020} {\bibfield
  {journal} {\bibinfo  {journal} {Commun. Math. Phys.}\ }\textbf {\bibinfo
  {volume} {43}},\ \bibinfo {pages} {199--220} (\bibinfo {year} {1975})},\
  \bibinfo {note} {[Erratum: Commun.Math.Phys. 46, 206 (1976)]}\BibitemShut
  {NoStop}%
\bibitem [{\citenamefont {Preskill}(1992)}]{Preskill:1992tc}%
  \BibitemOpen
  \bibfield  {author} {\bibinfo {author} {\bibfnamefont {John}\ \bibnamefont
  {Preskill}},\ }\bibfield  {title} {\enquote {\bibinfo {title} {{Do black
  holes destroy information?}}}\ }in\ \href@noop {} {\emph {\bibinfo
  {booktitle} {{International Symposium on Black holes, Membranes, Wormholes
  and Superstrings}}}}\ (\bibinfo {year} {1992})\ \Eprint
  {http://arxiv.org/abs/hep-th/9209058} {arXiv:hep-th/9209058} \BibitemShut
  {NoStop}%
\bibitem [{\citenamefont {Dvali}(2018)}]{Dvali:2018xpy}%
  \BibitemOpen
  \bibfield  {author} {\bibinfo {author} {\bibfnamefont {Gia}\ \bibnamefont
  {Dvali}},\ }\bibfield  {title} {\enquote {\bibinfo {title} {{A Microscopic
  Model of Holography: Survival by the Burden of Memory}},}\ }\href@noop {} {\
  (\bibinfo {year} {2018})},\ \Eprint {http://arxiv.org/abs/1810.02336}
  {arXiv:1810.02336 [hep-th]} \BibitemShut {NoStop}%
\bibitem [{\citenamefont {Dvali}\ \emph {et~al.}(2024)\citenamefont {Dvali},
  \citenamefont {Valbuena-Berm\'udez},\ and\ \citenamefont
  {Zantedeschi}}]{Dvali:2024hsb}%
  \BibitemOpen
  \bibfield  {author} {\bibinfo {author} {\bibfnamefont {Gia}\ \bibnamefont
  {Dvali}}, \bibinfo {author} {\bibfnamefont {Juan~Sebasti\'an}\ \bibnamefont
  {Valbuena-Berm\'udez}}, \ and\ \bibinfo {author} {\bibfnamefont {Michael}\
  \bibnamefont {Zantedeschi}},\ }\bibfield  {title} {\enquote {\bibinfo {title}
  {{Memory burden effect in black holes and solitons: Implications for PBH}},}\
  }\href {\doibase 10.1103/PhysRevD.110.056029} {\bibfield  {journal} {\bibinfo
   {journal} {Phys. Rev. D}\ }\textbf {\bibinfo {volume} {110}},\ \bibinfo
  {pages} {056029} (\bibinfo {year} {2024})},\ \Eprint
  {http://arxiv.org/abs/2405.13117} {arXiv:2405.13117 [hep-th]} \BibitemShut
  {NoStop}%
\bibitem [{\citenamefont {Balaji}\ \emph {et~al.}(2024)\citenamefont {Balaji},
  \citenamefont {Dom\`enech}, \citenamefont {Franciolini}, \citenamefont
  {Ganz},\ and\ \citenamefont {Tr\"ankle}}]{Balaji:2024hpu}%
  \BibitemOpen
  \bibfield  {author} {\bibinfo {author} {\bibfnamefont {Shyam}\ \bibnamefont
  {Balaji}}, \bibinfo {author} {\bibfnamefont {Guillem}\ \bibnamefont
  {Dom\`enech}}, \bibinfo {author} {\bibfnamefont {Gabriele}\ \bibnamefont
  {Franciolini}}, \bibinfo {author} {\bibfnamefont {Alexander}\ \bibnamefont
  {Ganz}}, \ and\ \bibinfo {author} {\bibfnamefont {Jan}\ \bibnamefont
  {Tr\"ankle}},\ }\bibfield  {title} {\enquote {\bibinfo {title} {{Probing
  modified Hawking evaporation with gravitational waves from the primordial
  black hole dominated universe}},}\ }\href@noop {} {\  (\bibinfo {year}
  {2024})},\ \Eprint {http://arxiv.org/abs/2403.14309} {arXiv:2403.14309
  [gr-qc]} \BibitemShut {NoStop}%
\bibitem [{\citenamefont {Barman}\ \emph
  {et~al.}(2024{\natexlab{a}})\citenamefont {Barman}, \citenamefont {Haque},\
  and\ \citenamefont {Zapata}}]{Barman:2024iht}%
  \BibitemOpen
  \bibfield  {author} {\bibinfo {author} {\bibfnamefont {Basabendu}\
  \bibnamefont {Barman}}, \bibinfo {author} {\bibfnamefont {Md~Riajul}\
  \bibnamefont {Haque}}, \ and\ \bibinfo {author} {\bibfnamefont {\'Oscar}\
  \bibnamefont {Zapata}},\ }\bibfield  {title} {\enquote {\bibinfo {title}
  {{Gravitational wave signatures of cogenesis from a burdened PBH}},}\ }\href
  {\doibase 10.1088/1475-7516/2024/09/020} {\bibfield  {journal} {\bibinfo
  {journal} {JCAP}\ }\textbf {\bibinfo {volume} {09}},\ \bibinfo {pages} {020}
  (\bibinfo {year} {2024}{\natexlab{a}})},\ \Eprint
  {http://arxiv.org/abs/2405.15858} {arXiv:2405.15858 [astro-ph.CO]}
  \BibitemShut {NoStop}%
\bibitem [{\citenamefont {Bhaumik}\ \emph {et~al.}(2024)\citenamefont
  {Bhaumik}, \citenamefont {Haque}, \citenamefont {Jain},\ and\ \citenamefont
  {Lewicki}}]{Bhaumik:2024qzd}%
  \BibitemOpen
  \bibfield  {author} {\bibinfo {author} {\bibfnamefont {Nilanjandev}\
  \bibnamefont {Bhaumik}}, \bibinfo {author} {\bibfnamefont {Md~Riajul}\
  \bibnamefont {Haque}}, \bibinfo {author} {\bibfnamefont {Rajeev~Kumar}\
  \bibnamefont {Jain}}, \ and\ \bibinfo {author} {\bibfnamefont {Marek}\
  \bibnamefont {Lewicki}},\ }\bibfield  {title} {\enquote {\bibinfo {title}
  {{Memory burden effect mimics reheating signatures on SGWB from ultra-low
  mass PBH domination}},}\ }\href@noop {} {\  (\bibinfo {year} {2024})},\
  \Eprint {http://arxiv.org/abs/2409.04436} {arXiv:2409.04436 [astro-ph.CO]}
  \BibitemShut {NoStop}%
\bibitem [{\citenamefont {Barman}\ \emph
  {et~al.}(2024{\natexlab{b}})\citenamefont {Barman}, \citenamefont {Loho},\
  and\ \citenamefont {Zapata}}]{Barman:2024ufm}%
  \BibitemOpen
  \bibfield  {author} {\bibinfo {author} {\bibfnamefont {Basabendu}\
  \bibnamefont {Barman}}, \bibinfo {author} {\bibfnamefont {Kousik}\
  \bibnamefont {Loho}}, \ and\ \bibinfo {author} {\bibfnamefont {\'Oscar}\
  \bibnamefont {Zapata}},\ }\bibfield  {title} {\enquote {\bibinfo {title}
  {{Constraining burdened PBHs with gravitational waves}},}\ }\href@noop {} {\
  (\bibinfo {year} {2024}{\natexlab{b}})},\ \Eprint
  {http://arxiv.org/abs/2409.05953} {arXiv:2409.05953 [gr-qc]} \BibitemShut
  {NoStop}%
\bibitem [{\citenamefont {Kohri}\ \emph {et~al.}(2024)\citenamefont {Kohri},
  \citenamefont {Terada},\ and\ \citenamefont {Yanagida}}]{Kohri:2024qpd}%
  \BibitemOpen
  \bibfield  {author} {\bibinfo {author} {\bibfnamefont {Kazunori}\
  \bibnamefont {Kohri}}, \bibinfo {author} {\bibfnamefont {Takahiro}\
  \bibnamefont {Terada}}, \ and\ \bibinfo {author} {\bibfnamefont {Tsutomu~T.}\
  \bibnamefont {Yanagida}},\ }\bibfield  {title} {\enquote {\bibinfo {title}
  {{Induced Gravitational Waves probing Primordial Black Hole Dark Matter with
  Memory Burden}},}\ }\href@noop {} {\  (\bibinfo {year} {2024})},\ \Eprint
  {http://arxiv.org/abs/2409.06365} {arXiv:2409.06365 [astro-ph.CO]}
  \BibitemShut {NoStop}%
\bibitem [{\citenamefont {Jiang}\ \emph {et~al.}(2024)\citenamefont {Jiang},
  \citenamefont {Yuan}, \citenamefont {Li},\ and\ \citenamefont
  {Huang}}]{Jiang:2024aju}%
  \BibitemOpen
  \bibfield  {author} {\bibinfo {author} {\bibfnamefont {Yang}\ \bibnamefont
  {Jiang}}, \bibinfo {author} {\bibfnamefont {Chen}\ \bibnamefont {Yuan}},
  \bibinfo {author} {\bibfnamefont {Chong-Zhi}\ \bibnamefont {Li}}, \ and\
  \bibinfo {author} {\bibfnamefont {Qing-Guo}\ \bibnamefont {Huang}},\
  }\bibfield  {title} {\enquote {\bibinfo {title} {{Constraints on the
  Primordial Black Hole Abundance through Scalar-Induced Gravitational Waves
  from Advanced LIGO and Virgo's First Three Observing Runs}},}\ }\href@noop {}
  {\  (\bibinfo {year} {2024})},\ \Eprint {http://arxiv.org/abs/2409.07976}
  {arXiv:2409.07976 [astro-ph.CO]} \BibitemShut {NoStop}%
\bibitem [{\citenamefont {Dvali}\ \emph {et~al.}(2020)\citenamefont {Dvali},
  \citenamefont {Eisemann}, \citenamefont {Michel},\ and\ \citenamefont
  {Zell}}]{Dvali:2020wft}%
  \BibitemOpen
  \bibfield  {author} {\bibinfo {author} {\bibfnamefont {Gia}\ \bibnamefont
  {Dvali}}, \bibinfo {author} {\bibfnamefont {Lukas}\ \bibnamefont {Eisemann}},
  \bibinfo {author} {\bibfnamefont {Marco}\ \bibnamefont {Michel}}, \ and\
  \bibinfo {author} {\bibfnamefont {Sebastian}\ \bibnamefont {Zell}},\
  }\bibfield  {title} {\enquote {\bibinfo {title} {{Black hole metamorphosis
  and stabilization by memory burden}},}\ }\href {\doibase
  10.1103/PhysRevD.102.103523} {\bibfield  {journal} {\bibinfo  {journal}
  {Phys. Rev. D}\ }\textbf {\bibinfo {volume} {102}},\ \bibinfo {pages}
  {103523} (\bibinfo {year} {2020})},\ \Eprint
  {http://arxiv.org/abs/2006.00011} {arXiv:2006.00011 [hep-th]} \BibitemShut
  {NoStop}%
\bibitem [{\citenamefont {Dvali}\ \emph {et~al.}(2021)\citenamefont {Dvali},
  \citenamefont {K\"uhnel},\ and\ \citenamefont {Zantedeschi}}]{Dvali:2021byy}%
  \BibitemOpen
  \bibfield  {author} {\bibinfo {author} {\bibfnamefont {Gia}\ \bibnamefont
  {Dvali}}, \bibinfo {author} {\bibfnamefont {Florian}\ \bibnamefont
  {K\"uhnel}}, \ and\ \bibinfo {author} {\bibfnamefont {Michael}\ \bibnamefont
  {Zantedeschi}},\ }\bibfield  {title} {\enquote {\bibinfo {title} {{Primordial
  black holes from confinement}},}\ }\href {\doibase
  10.1103/PhysRevD.104.123507} {\bibfield  {journal} {\bibinfo  {journal}
  {Phys. Rev. D}\ }\textbf {\bibinfo {volume} {104}},\ \bibinfo {pages}
  {123507} (\bibinfo {year} {2021})},\ \Eprint
  {http://arxiv.org/abs/2108.09471} {arXiv:2108.09471 [hep-ph]} \BibitemShut
  {NoStop}%
\bibitem [{\citenamefont {Alexandre}\ \emph {et~al.}(2024)\citenamefont
  {Alexandre}, \citenamefont {Dvali},\ and\ \citenamefont
  {Koutsangelas}}]{Alexandre:2024nuo}%
  \BibitemOpen
  \bibfield  {author} {\bibinfo {author} {\bibfnamefont {Ana}\ \bibnamefont
  {Alexandre}}, \bibinfo {author} {\bibfnamefont {Gia}\ \bibnamefont {Dvali}},
  \ and\ \bibinfo {author} {\bibfnamefont {Emmanouil}\ \bibnamefont
  {Koutsangelas}},\ }\bibfield  {title} {\enquote {\bibinfo {title} {{New mass
  window for primordial black holes as dark matter from the memory burden
  effect}},}\ }\href {\doibase 10.1103/PhysRevD.110.036004} {\bibfield
  {journal} {\bibinfo  {journal} {Phys. Rev. D}\ }\textbf {\bibinfo {volume}
  {110}},\ \bibinfo {pages} {036004} (\bibinfo {year} {2024})},\ \Eprint
  {http://arxiv.org/abs/2402.14069} {arXiv:2402.14069 [hep-ph]} \BibitemShut
  {NoStop}%
\bibitem [{\citenamefont {Thoss}\ \emph {et~al.}(2024)\citenamefont {Thoss},
  \citenamefont {Burkert},\ and\ \citenamefont {Kohri}}]{Thoss:2024hsr}%
  \BibitemOpen
  \bibfield  {author} {\bibinfo {author} {\bibfnamefont {Valentin}\
  \bibnamefont {Thoss}}, \bibinfo {author} {\bibfnamefont {Andreas}\
  \bibnamefont {Burkert}}, \ and\ \bibinfo {author} {\bibfnamefont {Kazunori}\
  \bibnamefont {Kohri}},\ }\bibfield  {title} {\enquote {\bibinfo {title}
  {{Breakdown of hawking evaporation opens new mass window for primordial black
  holes as dark matter candidate}},}\ }\href {\doibase 10.1093/mnras/stae1098}
  {\bibfield  {journal} {\bibinfo  {journal} {Mon. Not. Roy. Astron. Soc.}\
  }\textbf {\bibinfo {volume} {532}},\ \bibinfo {pages} {451--459} (\bibinfo
  {year} {2024})},\ \Eprint {http://arxiv.org/abs/2402.17823} {arXiv:2402.17823
  [astro-ph.CO]} \BibitemShut {NoStop}%
\bibitem [{\citenamefont {Haque}\ \emph {et~al.}(2024)\citenamefont {Haque},
  \citenamefont {Maity}, \citenamefont {Maity},\ and\ \citenamefont
  {Mambrini}}]{Haque:2024eyh}%
  \BibitemOpen
  \bibfield  {author} {\bibinfo {author} {\bibfnamefont {Md~Riajul}\
  \bibnamefont {Haque}}, \bibinfo {author} {\bibfnamefont {Suvashis}\
  \bibnamefont {Maity}}, \bibinfo {author} {\bibfnamefont {Debaprasad}\
  \bibnamefont {Maity}}, \ and\ \bibinfo {author} {\bibfnamefont {Yann}\
  \bibnamefont {Mambrini}},\ }\bibfield  {title} {\enquote {\bibinfo {title}
  {{Quantum effects on the evaporation of PBHs: contributions to dark
  matter}},}\ }\href {\doibase 10.1088/1475-7516/2024/07/002} {\bibfield
  {journal} {\bibinfo  {journal} {JCAP}\ }\textbf {\bibinfo {volume} {07}},\
  \bibinfo {pages} {002} (\bibinfo {year} {2024})},\ \Eprint
  {http://arxiv.org/abs/2404.16815} {arXiv:2404.16815 [hep-ph]} \BibitemShut
  {NoStop}%
\bibitem [{\citenamefont {Chianese}\ \emph {et~al.}(2024)\citenamefont
  {Chianese}, \citenamefont {Boccia}, \citenamefont {Iocco}, \citenamefont
  {Miele},\ and\ \citenamefont {Saviano}}]{Chianese:2024rsn}%
  \BibitemOpen
  \bibfield  {author} {\bibinfo {author} {\bibfnamefont {Marco}\ \bibnamefont
  {Chianese}}, \bibinfo {author} {\bibfnamefont {Andrea}\ \bibnamefont
  {Boccia}}, \bibinfo {author} {\bibfnamefont {Fabio}\ \bibnamefont {Iocco}},
  \bibinfo {author} {\bibfnamefont {Gennaro}\ \bibnamefont {Miele}}, \ and\
  \bibinfo {author} {\bibfnamefont {Ninetta}\ \bibnamefont {Saviano}},\
  }\bibfield  {title} {\enquote {\bibinfo {title} {{The light burden of memory:
  constraining primordial black holes with high-energy neutrinos}},}\
  }\href@noop {} {\  (\bibinfo {year} {2024})},\ \Eprint
  {http://arxiv.org/abs/2410.07604} {arXiv:2410.07604 [astro-ph.HE]}
  \BibitemShut {NoStop}%
\bibitem [{\citenamefont {Zantedeschi}\ and\ \citenamefont
  {Visinelli}(2024)}]{Zantedeschi:2024ram}%
  \BibitemOpen
  \bibfield  {author} {\bibinfo {author} {\bibfnamefont {Michael}\ \bibnamefont
  {Zantedeschi}}\ and\ \bibinfo {author} {\bibfnamefont {Luca}\ \bibnamefont
  {Visinelli}},\ }\bibfield  {title} {\enquote {\bibinfo {title}
  {{Memory-Burdened Primordial Black Holes as Astrophysical Particle
  Accelerators}},}\ }\href@noop {} {\  (\bibinfo {year} {2024})},\ \Eprint
  {http://arxiv.org/abs/2410.07037} {arXiv:2410.07037 [astro-ph.HE]}
  \BibitemShut {NoStop}%
\bibitem [{\citenamefont {Barman}\ \emph
  {et~al.}(2024{\natexlab{c}})\citenamefont {Barman}, \citenamefont {Loho},\
  and\ \citenamefont {Zapata}}]{Barman:2024kfj}%
  \BibitemOpen
  \bibfield  {author} {\bibinfo {author} {\bibfnamefont {Basabendu}\
  \bibnamefont {Barman}}, \bibinfo {author} {\bibfnamefont {Kousik}\
  \bibnamefont {Loho}}, \ and\ \bibinfo {author} {\bibfnamefont {\'Oscar}\
  \bibnamefont {Zapata}},\ }\bibfield  {title} {\enquote {\bibinfo {title}
  {{Asymmetries from a charged memory-burdened PBH}},}\ }\href@noop {} {\
  (\bibinfo {year} {2024}{\natexlab{c}})},\ \Eprint
  {http://arxiv.org/abs/2412.13254} {arXiv:2412.13254 [hep-ph]} \BibitemShut
  {NoStop}%
\bibitem [{\citenamefont {Borah}\ and\ \citenamefont
  {Das}(2024)}]{Borah:2024bcr}%
  \BibitemOpen
  \bibfield  {author} {\bibinfo {author} {\bibfnamefont {Debasish}\
  \bibnamefont {Borah}}\ and\ \bibinfo {author} {\bibfnamefont {Nayan}\
  \bibnamefont {Das}},\ }\bibfield  {title} {\enquote {\bibinfo {title}
  {{Successful cogenesis of baryon and dark matter from memory-burdened
  PBH}},}\ }\href@noop {} {\  (\bibinfo {year} {2024})},\ \Eprint
  {http://arxiv.org/abs/2410.16403} {arXiv:2410.16403 [hep-ph]} \BibitemShut
  {NoStop}%
\bibitem [{\citenamefont {Athron}\ \emph {et~al.}(2024)\citenamefont {Athron},
  \citenamefont {Chianese}, \citenamefont {Datta}, \citenamefont {Samanta},\
  and\ \citenamefont {Saviano}}]{Athron:2024fcj}%
  \BibitemOpen
  \bibfield  {author} {\bibinfo {author} {\bibfnamefont {Peter}\ \bibnamefont
  {Athron}}, \bibinfo {author} {\bibfnamefont {Marco}\ \bibnamefont
  {Chianese}}, \bibinfo {author} {\bibfnamefont {Satyabrata}\ \bibnamefont
  {Datta}}, \bibinfo {author} {\bibfnamefont {Rome}\ \bibnamefont {Samanta}}, \
  and\ \bibinfo {author} {\bibfnamefont {Ninetta}\ \bibnamefont {Saviano}},\
  }\bibfield  {title} {\enquote {\bibinfo {title} {{Impact of memory-burdened
  black holes on primordial gravitational waves in light of Pulsar Timing
  Array}},}\ }\href@noop {} {\  (\bibinfo {year} {2024})},\ \Eprint
  {http://arxiv.org/abs/2411.19286} {arXiv:2411.19286 [astro-ph.CO]}
  \BibitemShut {NoStop}%
\bibitem [{\citenamefont {Loc}(2025)}]{Loc:2024qbz}%
  \BibitemOpen
  \bibfield  {author} {\bibinfo {author} {\bibfnamefont {Ngo Phuc~Duc}\
  \bibnamefont {Loc}},\ }\bibfield  {title} {\enquote {\bibinfo {title}
  {{Gravitational waves from burdened primordial black holes dark matter}},}\
  }\href {\doibase 10.1103/PhysRevD.111.023509} {\bibfield  {journal} {\bibinfo
   {journal} {Phys. Rev. D}\ }\textbf {\bibinfo {volume} {111}},\ \bibinfo
  {pages} {023509} (\bibinfo {year} {2025})},\ \Eprint
  {http://arxiv.org/abs/2410.17544} {arXiv:2410.17544 [gr-qc]} \BibitemShut
  {NoStop}%
\bibitem [{\citenamefont {Bandyopadhyay}\ \emph {et~al.}(2025)\citenamefont
  {Bandyopadhyay}, \citenamefont {Borah},\ and\ \citenamefont
  {Das}}]{Bandyopadhyay:2025ast}%
  \BibitemOpen
  \bibfield  {author} {\bibinfo {author} {\bibfnamefont {Disha}\ \bibnamefont
  {Bandyopadhyay}}, \bibinfo {author} {\bibfnamefont {Debasish}\ \bibnamefont
  {Borah}}, \ and\ \bibinfo {author} {\bibfnamefont {Nayan}\ \bibnamefont
  {Das}},\ }\bibfield  {title} {\enquote {\bibinfo {title} {{Axion misalignment
  with memory-burdened PBH}},}\ }\href@noop {} {\  (\bibinfo {year} {2025})},\
  \Eprint {http://arxiv.org/abs/2501.04076} {arXiv:2501.04076 [hep-ph]}
  \BibitemShut {NoStop}%
\bibitem [{\citenamefont {Fukugita}\ and\ \citenamefont
  {Yanagida}(1986)}]{fukugita1986barygenesis}%
  \BibitemOpen
  \bibfield  {author} {\bibinfo {author} {\bibfnamefont {Masataka}\
  \bibnamefont {Fukugita}}\ and\ \bibinfo {author} {\bibfnamefont {Tsutomu}\
  \bibnamefont {Yanagida}},\ }\bibfield  {title} {\enquote {\bibinfo {title}
  {Barygenesis without grand unification},}\ }\href@noop {} {\bibfield
  {journal} {\bibinfo  {journal} {Physics Letters B}\ }\textbf {\bibinfo
  {volume} {174}},\ \bibinfo {pages} {45--47} (\bibinfo {year}
  {1986})}\BibitemShut {NoStop}%
\bibitem [{\citenamefont {Davidson}\ \emph {et~al.}(2008)\citenamefont
  {Davidson}, \citenamefont {Nardi},\ and\ \citenamefont
  {Nir}}]{Davidson:2008bu}%
  \BibitemOpen
  \bibfield  {author} {\bibinfo {author} {\bibfnamefont {Sacha}\ \bibnamefont
  {Davidson}}, \bibinfo {author} {\bibfnamefont {Enrico}\ \bibnamefont
  {Nardi}}, \ and\ \bibinfo {author} {\bibfnamefont {Yosef}\ \bibnamefont
  {Nir}},\ }\bibfield  {title} {\enquote {\bibinfo {title} {{Leptogenesis}},}\
  }\href {\doibase 10.1016/j.physrep.2008.06.002} {\bibfield  {journal}
  {\bibinfo  {journal} {Phys. Rept.}\ }\textbf {\bibinfo {volume} {466}},\
  \bibinfo {pages} {105--177} (\bibinfo {year} {2008})},\ \Eprint
  {http://arxiv.org/abs/0802.2962} {arXiv:0802.2962 [hep-ph]} \BibitemShut
  {NoStop}%
\bibitem [{\citenamefont {Fong}\ \emph {et~al.}(2012)\citenamefont {Fong},
  \citenamefont {Nardi},\ and\ \citenamefont {Riotto}}]{Fong:2012buy}%
  \BibitemOpen
  \bibfield  {author} {\bibinfo {author} {\bibfnamefont {Chee~Sheng}\
  \bibnamefont {Fong}}, \bibinfo {author} {\bibfnamefont {Enrico}\ \bibnamefont
  {Nardi}}, \ and\ \bibinfo {author} {\bibfnamefont {Antonio}\ \bibnamefont
  {Riotto}},\ }\bibfield  {title} {\enquote {\bibinfo {title} {{Leptogenesis in
  the Universe}},}\ }\href {\doibase 10.1155/2012/158303} {\bibfield  {journal}
  {\bibinfo  {journal} {Adv. High Energy Phys.}\ }\textbf {\bibinfo {volume}
  {2012}},\ \bibinfo {pages} {158303} (\bibinfo {year} {2012})},\ \Eprint
  {http://arxiv.org/abs/1301.3062} {arXiv:1301.3062 [hep-ph]} \BibitemShut
  {NoStop}%
\bibitem [{\citenamefont {Buchmuller}\ \emph
  {et~al.}(2005{\natexlab{a}})\citenamefont {Buchmuller}, \citenamefont
  {Peccei},\ and\ \citenamefont {Yanagida}}]{Buchmuller:2005eh}%
  \BibitemOpen
  \bibfield  {author} {\bibinfo {author} {\bibfnamefont {W.}~\bibnamefont
  {Buchmuller}}, \bibinfo {author} {\bibfnamefont {R.~D.}\ \bibnamefont
  {Peccei}}, \ and\ \bibinfo {author} {\bibfnamefont {T.}~\bibnamefont
  {Yanagida}},\ }\bibfield  {title} {\enquote {\bibinfo {title} {{Leptogenesis
  as the origin of matter}},}\ }\href {\doibase
  10.1146/annurev.nucl.55.090704.151558} {\bibfield  {journal} {\bibinfo
  {journal} {Ann. Rev. Nucl. Part. Sci.}\ }\textbf {\bibinfo {volume} {55}},\
  \bibinfo {pages} {311--355} (\bibinfo {year} {2005}{\natexlab{a}})},\ \Eprint
  {http://arxiv.org/abs/hep-ph/0502169} {arXiv:hep-ph/0502169} \BibitemShut
  {NoStop}%
\bibitem [{\citenamefont {Di~Bari}(2012)}]{DiBari:2012fz}%
  \BibitemOpen
  \bibfield  {author} {\bibinfo {author} {\bibfnamefont {Pasquale}\
  \bibnamefont {Di~Bari}},\ }\bibfield  {title} {\enquote {\bibinfo {title}
  {{An introduction to leptogenesis and neutrino properties}},}\ }\href
  {\doibase 10.1080/00107514.2012.701096} {\bibfield  {journal} {\bibinfo
  {journal} {Contemp. Phys.}\ }\textbf {\bibinfo {volume} {53}},\ \bibinfo
  {pages} {315--338} (\bibinfo {year} {2012})},\ \Eprint
  {http://arxiv.org/abs/1206.3168} {arXiv:1206.3168 [hep-ph]} \BibitemShut
  {NoStop}%
\bibitem [{\citenamefont {Harvey}\ and\ \citenamefont
  {Turner}(1990)}]{Harvey:1990qw}%
  \BibitemOpen
  \bibfield  {author} {\bibinfo {author} {\bibfnamefont {Jeffrey~A.}\
  \bibnamefont {Harvey}}\ and\ \bibinfo {author} {\bibfnamefont {Michael~S.}\
  \bibnamefont {Turner}},\ }\bibfield  {title} {\enquote {\bibinfo {title}
  {{Cosmological baryon and lepton number in the presence of electroweak
  fermion number violation}},}\ }\href {\doibase 10.1103/PhysRevD.42.3344}
  {\bibfield  {journal} {\bibinfo  {journal} {Phys. Rev. D}\ }\textbf {\bibinfo
  {volume} {42}},\ \bibinfo {pages} {3344--3349} (\bibinfo {year}
  {1990})}\BibitemShut {NoStop}%
\bibitem [{\citenamefont {Kuzmin}\ \emph {et~al.}(1985)\citenamefont {Kuzmin},
  \citenamefont {Rubakov},\ and\ \citenamefont {Shaposhnikov}}]{Kuzmin:1985mm}%
  \BibitemOpen
  \bibfield  {author} {\bibinfo {author} {\bibfnamefont {V.~A.}\ \bibnamefont
  {Kuzmin}}, \bibinfo {author} {\bibfnamefont {V.~A.}\ \bibnamefont {Rubakov}},
  \ and\ \bibinfo {author} {\bibfnamefont {M.~E.}\ \bibnamefont
  {Shaposhnikov}},\ }\bibfield  {title} {\enquote {\bibinfo {title} {{On the
  Anomalous Electroweak Baryon Number Nonconservation in the Early
  Universe}},}\ }\href {\doibase 10.1016/0370-2693(85)91028-7} {\bibfield
  {journal} {\bibinfo  {journal} {Phys. Lett. B}\ }\textbf {\bibinfo {volume}
  {155}},\ \bibinfo {pages} {36} (\bibinfo {year} {1985})}\BibitemShut
  {NoStop}%
\bibitem [{\citenamefont {Khlebnikov}\ and\ \citenamefont
  {Shaposhnikov}(1988)}]{Khlebnikov:1988sr}%
  \BibitemOpen
  \bibfield  {author} {\bibinfo {author} {\bibfnamefont {S.~Yu.}\ \bibnamefont
  {Khlebnikov}}\ and\ \bibinfo {author} {\bibfnamefont {M.~E.}\ \bibnamefont
  {Shaposhnikov}},\ }\bibfield  {title} {\enquote {\bibinfo {title} {{The
  Statistical Theory of Anomalous Fermion Number Nonconservation}},}\ }\href
  {\doibase 10.1016/0550-3213(88)90133-2} {\bibfield  {journal} {\bibinfo
  {journal} {Nucl. Phys. B}\ }\textbf {\bibinfo {volume} {308}},\ \bibinfo
  {pages} {885--912} (\bibinfo {year} {1988})}\BibitemShut {NoStop}%
\bibitem [{\citenamefont {Perez-Gonzalez}\ and\ \citenamefont
  {Turner}(2021)}]{Perez-Gonzalez:2020vnz}%
  \BibitemOpen
  \bibfield  {author} {\bibinfo {author} {\bibfnamefont {Yuber~F.}\
  \bibnamefont {Perez-Gonzalez}}\ and\ \bibinfo {author} {\bibfnamefont
  {Jessica}\ \bibnamefont {Turner}},\ }\bibfield  {title} {\enquote {\bibinfo
  {title} {{Assessing the tension between a black hole dominated early universe
  and leptogenesis}},}\ }\href {\doibase 10.1103/PhysRevD.104.103021}
  {\bibfield  {journal} {\bibinfo  {journal} {Phys. Rev. D}\ }\textbf {\bibinfo
  {volume} {104}},\ \bibinfo {pages} {103021} (\bibinfo {year} {2021})},\
  \Eprint {http://arxiv.org/abs/2010.03565} {arXiv:2010.03565 [hep-ph]}
  \BibitemShut {NoStop}%
\bibitem [{\citenamefont {Jyoti~Das}\ \emph {et~al.}(2021)\citenamefont
  {Jyoti~Das}, \citenamefont {Mahanta},\ and\ \citenamefont
  {Borah}}]{JyotiDas:2021shi}%
  \BibitemOpen
  \bibfield  {author} {\bibinfo {author} {\bibfnamefont {Suruj}\ \bibnamefont
  {Jyoti~Das}}, \bibinfo {author} {\bibfnamefont {Devabrat}\ \bibnamefont
  {Mahanta}}, \ and\ \bibinfo {author} {\bibfnamefont {Debasish}\ \bibnamefont
  {Borah}},\ }\bibfield  {title} {\enquote {\bibinfo {title} {{Low scale
  leptogenesis and dark matter in the presence of primordial black holes}},}\
  }\href {\doibase 10.1088/1475-7516/2021/11/019} {\bibfield  {journal}
  {\bibinfo  {journal} {JCAP}\ }\textbf {\bibinfo {volume} {11}},\ \bibinfo
  {pages} {019} (\bibinfo {year} {2021})},\ \Eprint
  {http://arxiv.org/abs/2104.14496} {arXiv:2104.14496 [hep-ph]} \BibitemShut
  {NoStop}%
\bibitem [{\citenamefont {Bernal}\ \emph
  {et~al.}(2022{\natexlab{b}})\citenamefont {Bernal}, \citenamefont {Fong},
  \citenamefont {Perez-Gonzalez},\ and\ \citenamefont
  {Turner}}]{Bernal:2022pue}%
  \BibitemOpen
  \bibfield  {author} {\bibinfo {author} {\bibfnamefont {Nicol\'as}\
  \bibnamefont {Bernal}}, \bibinfo {author} {\bibfnamefont {Chee~Sheng}\
  \bibnamefont {Fong}}, \bibinfo {author} {\bibfnamefont {Yuber~F.}\
  \bibnamefont {Perez-Gonzalez}}, \ and\ \bibinfo {author} {\bibfnamefont
  {Jessica}\ \bibnamefont {Turner}},\ }\bibfield  {title} {\enquote {\bibinfo
  {title} {{Rescuing high-scale leptogenesis using primordial black holes}},}\
  }\href {\doibase 10.1103/PhysRevD.106.035019} {\bibfield  {journal} {\bibinfo
   {journal} {Phys. Rev. D}\ }\textbf {\bibinfo {volume} {106}},\ \bibinfo
  {pages} {035019} (\bibinfo {year} {2022}{\natexlab{b}})},\ \Eprint
  {http://arxiv.org/abs/2203.08823} {arXiv:2203.08823 [hep-ph]} \BibitemShut
  {NoStop}%
\bibitem [{\citenamefont {Calabrese}\ \emph {et~al.}(2023)\citenamefont
  {Calabrese}, \citenamefont {Chianese}, \citenamefont {Gunn}, \citenamefont
  {Miele}, \citenamefont {Morisi},\ and\ \citenamefont
  {Saviano}}]{Calabrese:2023key}%
  \BibitemOpen
  \bibfield  {author} {\bibinfo {author} {\bibfnamefont {Roberta}\ \bibnamefont
  {Calabrese}}, \bibinfo {author} {\bibfnamefont {Marco}\ \bibnamefont
  {Chianese}}, \bibinfo {author} {\bibfnamefont {Jacob}\ \bibnamefont {Gunn}},
  \bibinfo {author} {\bibfnamefont {Gennaro}\ \bibnamefont {Miele}}, \bibinfo
  {author} {\bibfnamefont {Stefano}\ \bibnamefont {Morisi}}, \ and\ \bibinfo
  {author} {\bibfnamefont {Ninetta}\ \bibnamefont {Saviano}},\ }\bibfield
  {title} {\enquote {\bibinfo {title} {{Limits on light primordial black holes
  from high-scale leptogenesis}},}\ }\href {\doibase
  10.1103/PhysRevD.107.123537} {\bibfield  {journal} {\bibinfo  {journal}
  {Phys. Rev. D}\ }\textbf {\bibinfo {volume} {107}},\ \bibinfo {pages}
  {123537} (\bibinfo {year} {2023})},\ \Eprint
  {http://arxiv.org/abs/2305.13369} {arXiv:2305.13369 [hep-ph]} \BibitemShut
  {NoStop}%
\bibitem [{\citenamefont {Calabrese}\ \emph {et~al.}(2024)\citenamefont
  {Calabrese}, \citenamefont {Chianese}, \citenamefont {Gunn}, \citenamefont
  {Miele}, \citenamefont {Morisi},\ and\ \citenamefont
  {Saviano}}]{Calabrese:2023bxz}%
  \BibitemOpen
  \bibfield  {author} {\bibinfo {author} {\bibfnamefont {Roberta}\ \bibnamefont
  {Calabrese}}, \bibinfo {author} {\bibfnamefont {Marco}\ \bibnamefont
  {Chianese}}, \bibinfo {author} {\bibfnamefont {Jacob}\ \bibnamefont {Gunn}},
  \bibinfo {author} {\bibfnamefont {Gennaro}\ \bibnamefont {Miele}}, \bibinfo
  {author} {\bibfnamefont {Stefano}\ \bibnamefont {Morisi}}, \ and\ \bibinfo
  {author} {\bibfnamefont {Ninetta}\ \bibnamefont {Saviano}},\ }\bibfield
  {title} {\enquote {\bibinfo {title} {{Impact of primordial black holes on
  heavy neutral leptons searches in the framework of resonant leptogenesis}},}\
  }\href {\doibase 10.1103/PhysRevD.109.103001} {\bibfield  {journal} {\bibinfo
   {journal} {Phys. Rev. D}\ }\textbf {\bibinfo {volume} {109}},\ \bibinfo
  {pages} {103001} (\bibinfo {year} {2024})},\ \Eprint
  {http://arxiv.org/abs/2311.13276} {arXiv:2311.13276 [hep-ph]} \BibitemShut
  {NoStop}%
\bibitem [{\citenamefont {Gunn}\ \emph {et~al.}(2024)\citenamefont {Gunn},
  \citenamefont {Heurtier}, \citenamefont {Perez-Gonzalez},\ and\ \citenamefont
  {Turner}}]{Gunn:2024xaq}%
  \BibitemOpen
  \bibfield  {author} {\bibinfo {author} {\bibfnamefont {Jacob}\ \bibnamefont
  {Gunn}}, \bibinfo {author} {\bibfnamefont {Lucien}\ \bibnamefont {Heurtier}},
  \bibinfo {author} {\bibfnamefont {Yuber~F.}\ \bibnamefont {Perez-Gonzalez}},
  \ and\ \bibinfo {author} {\bibfnamefont {Jessica}\ \bibnamefont {Turner}},\
  }\bibfield  {title} {\enquote {\bibinfo {title} {{Primordial Black Hole Hot
  Spots and Out-of-Equilibrium Dynamics}},}\ }\href@noop {} {\  (\bibinfo
  {year} {2024})},\ \Eprint {http://arxiv.org/abs/2409.02173} {arXiv:2409.02173
  [hep-ph]} \BibitemShut {NoStop}%
\bibitem [{\citenamefont {Carr}(1975)}]{Carr:1975qj}%
  \BibitemOpen
  \bibfield  {author} {\bibinfo {author} {\bibfnamefont {Bernard~J.}\
  \bibnamefont {Carr}},\ }\bibfield  {title} {\enquote {\bibinfo {title} {{The
  Primordial black hole mass spectrum}},}\ }\href {\doibase 10.1086/153853}
  {\bibfield  {journal} {\bibinfo  {journal} {Astrophys. J.}\ }\textbf
  {\bibinfo {volume} {201}},\ \bibinfo {pages} {1--19} (\bibinfo {year}
  {1975})}\BibitemShut {NoStop}%
\bibitem [{\citenamefont {Akrami}\ \emph {et~al.}(2020)\citenamefont {Akrami}
  \emph {et~al.}}]{Planck:2018jri}%
  \BibitemOpen
  \bibfield  {author} {\bibinfo {author} {\bibfnamefont {Y.}~\bibnamefont
  {Akrami}} \emph {et~al.} (\bibinfo {collaboration} {Planck}),\ }\bibfield
  {title} {\enquote {\bibinfo {title} {{Planck 2018 results. X. Constraints on
  inflation}},}\ }\href {\doibase 10.1051/0004-6361/201833887} {\bibfield
  {journal} {\bibinfo  {journal} {Astron. Astrophys.}\ }\textbf {\bibinfo
  {volume} {641}},\ \bibinfo {pages} {A10} (\bibinfo {year} {2020})},\ \Eprint
  {http://arxiv.org/abs/1807.06211} {arXiv:1807.06211 [astro-ph.CO]}
  \BibitemShut {NoStop}%
\bibitem [{\citenamefont {Hawking}(1974)}]{Hawking:1974rv}%
  \BibitemOpen
  \bibfield  {author} {\bibinfo {author} {\bibfnamefont {S.~W.}\ \bibnamefont
  {Hawking}},\ }\bibfield  {title} {\enquote {\bibinfo {title} {{Black hole
  explosions}},}\ }\href {\doibase 10.1038/248030a0} {\bibfield  {journal}
  {\bibinfo  {journal} {Nature}\ }\textbf {\bibinfo {volume} {248}},\ \bibinfo
  {pages} {30--31} (\bibinfo {year} {1974})}\BibitemShut {NoStop}%
\bibitem [{\citenamefont {Datta}\ \emph {et~al.}(2021)\citenamefont {Datta},
  \citenamefont {Ghosal},\ and\ \citenamefont {Samanta}}]{Datta:2020bht}%
  \BibitemOpen
  \bibfield  {author} {\bibinfo {author} {\bibfnamefont {Satyabrata}\
  \bibnamefont {Datta}}, \bibinfo {author} {\bibfnamefont {Ambar}\ \bibnamefont
  {Ghosal}}, \ and\ \bibinfo {author} {\bibfnamefont {Rome}\ \bibnamefont
  {Samanta}},\ }\bibfield  {title} {\enquote {\bibinfo {title} {{Baryogenesis
  from ultralight primordial black holes and strong gravitational waves from
  cosmic strings}},}\ }\href {\doibase 10.1088/1475-7516/2021/08/021}
  {\bibfield  {journal} {\bibinfo  {journal} {JCAP}\ }\textbf {\bibinfo
  {volume} {08}},\ \bibinfo {pages} {021} (\bibinfo {year} {2021})},\ \Eprint
  {http://arxiv.org/abs/2012.14981} {arXiv:2012.14981 [hep-ph]} \BibitemShut
  {NoStop}%
\bibitem [{\citenamefont {Lunardini}\ and\ \citenamefont
  {Perez-Gonzalez}(2020)}]{Lunardini:2019zob}%
  \BibitemOpen
  \bibfield  {author} {\bibinfo {author} {\bibfnamefont {Cecilia}\ \bibnamefont
  {Lunardini}}\ and\ \bibinfo {author} {\bibfnamefont {Yuber~F.}\ \bibnamefont
  {Perez-Gonzalez}},\ }\bibfield  {title} {\enquote {\bibinfo {title} {{Dirac
  and Majorana neutrino signatures of primordial black holes}},}\ }\href
  {\doibase 10.1088/1475-7516/2020/08/014} {\bibfield  {journal} {\bibinfo
  {journal} {JCAP}\ }\textbf {\bibinfo {volume} {08}},\ \bibinfo {pages} {014}
  (\bibinfo {year} {2020})},\ \Eprint {http://arxiv.org/abs/1910.07864}
  {arXiv:1910.07864 [hep-ph]} \BibitemShut {NoStop}%
\bibitem [{\citenamefont {Yanagida}(1980)}]{Yanagida:1980xy}%
  \BibitemOpen
  \bibfield  {author} {\bibinfo {author} {\bibfnamefont {Tsutomu}\ \bibnamefont
  {Yanagida}},\ }\bibfield  {title} {\enquote {\bibinfo {title} {{Horizontal
  Symmetry and Masses of Neutrinos}},}\ }\href {\doibase 10.1143/PTP.64.1103}
  {\bibfield  {journal} {\bibinfo  {journal} {Prog. Theor. Phys.}\ }\textbf
  {\bibinfo {volume} {64}},\ \bibinfo {pages} {1103} (\bibinfo {year}
  {1980})}\BibitemShut {NoStop}%
\bibitem [{\citenamefont {Mohapatra}\ and\ \citenamefont
  {Senjanovi\ifmmode~\acute{c}\else
  \'{c}\fi{}}(1981)}]{MohapatraRabindraSenjanovi}%
  \BibitemOpen
  \bibfield  {author} {\bibinfo {author} {\bibfnamefont {Rabindra~N.}\
  \bibnamefont {Mohapatra}}\ and\ \bibinfo {author} {\bibfnamefont {Goran}\
  \bibnamefont {Senjanovi\ifmmode~\acute{c}\else \'{c}\fi{}}},\ }\bibfield
  {title} {\enquote {\bibinfo {title} {Neutrino masses and mixings in gauge
  models with spontaneous parity violation},}\ }\href {\doibase
  10.1103/PhysRevD.23.165} {\bibfield  {journal} {\bibinfo  {journal} {Phys.
  Rev. D}\ }\textbf {\bibinfo {volume} {23}},\ \bibinfo {pages} {165--180}
  (\bibinfo {year} {1981})}\BibitemShut {NoStop}%
\bibitem [{\citenamefont
  {Pontecorvo}(1957{\natexlab{a}})}]{pontecorvo1957mesonium}%
  \BibitemOpen
  \bibfield  {author} {\bibinfo {author} {\bibfnamefont {Bruno}\ \bibnamefont
  {Pontecorvo}},\ }\bibfield  {title} {\enquote {\bibinfo {title} {Mesonium and
  antimesonium},}\ }\href@noop {} {\bibfield  {journal} {\bibinfo  {journal}
  {Zhur. Eksptl'. i Teoret. Fiz.}\ }\textbf {\bibinfo {volume} {33}} (\bibinfo
  {year} {1957}{\natexlab{a}})}\BibitemShut {NoStop}%
\bibitem [{\citenamefont
  {Pontecorvo}(1957{\natexlab{b}})}]{pontecorvo1957inverse}%
  \BibitemOpen
  \bibfield  {author} {\bibinfo {author} {\bibfnamefont {Bruno}\ \bibnamefont
  {Pontecorvo}},\ }\bibfield  {title} {\enquote {\bibinfo {title} {Inverse beta
  processes and nonconservation of lepton charge},}\ }\href@noop {} {\bibfield
  {journal} {\bibinfo  {journal} {Zh. Eksp. Teor. Fiz}\ }\textbf {\bibinfo
  {volume} {34}},\ \bibinfo {pages} {247} (\bibinfo {year}
  {1957}{\natexlab{b}})}\BibitemShut {NoStop}%
\bibitem [{\citenamefont {Maki}\ \emph {et~al.}(1962)\citenamefont {Maki},
  \citenamefont {Nakagawa},\ and\ \citenamefont {Sakata}}]{Maki:1962mu}%
  \BibitemOpen
  \bibfield  {author} {\bibinfo {author} {\bibfnamefont {Ziro}\ \bibnamefont
  {Maki}}, \bibinfo {author} {\bibfnamefont {Masami}\ \bibnamefont {Nakagawa}},
  \ and\ \bibinfo {author} {\bibfnamefont {Shoichi}\ \bibnamefont {Sakata}},\
  }\bibfield  {title} {\enquote {\bibinfo {title} {{Remarks on the unified
  model of elementary particles}},}\ }\href {\doibase 10.1143/PTP.28.870}
  {\bibfield  {journal} {\bibinfo  {journal} {Prog. Theor. Phys.}\ }\textbf
  {\bibinfo {volume} {28}},\ \bibinfo {pages} {870--880} (\bibinfo {year}
  {1962})}\BibitemShut {NoStop}%
\bibitem [{\citenamefont {Pontecorvo}(1967)}]{Pontecorvo:1967fh}%
  \BibitemOpen
  \bibfield  {author} {\bibinfo {author} {\bibfnamefont {B.}~\bibnamefont
  {Pontecorvo}},\ }\bibfield  {title} {\enquote {\bibinfo {title} {{Neutrino
  Experiments and the Problem of Conservation of Leptonic Charge}},}\
  }\href@noop {} {\bibfield  {journal} {\bibinfo  {journal} {Zh. Eksp. Teor.
  Fiz.}\ }\textbf {\bibinfo {volume} {53}},\ \bibinfo {pages} {1717--1725}
  (\bibinfo {year} {1967})}\BibitemShut {NoStop}%
\bibitem [{\citenamefont {Gribov}\ and\ \citenamefont
  {Pontecorvo}(1969)}]{Gribov:1968kq}%
  \BibitemOpen
  \bibfield  {author} {\bibinfo {author} {\bibfnamefont {V.~N.}\ \bibnamefont
  {Gribov}}\ and\ \bibinfo {author} {\bibfnamefont {B.}~\bibnamefont
  {Pontecorvo}},\ }\bibfield  {title} {\enquote {\bibinfo {title} {{Neutrino
  astronomy and lepton charge}},}\ }\href {\doibase
  10.1016/0370-2693(69)90525-5} {\bibfield  {journal} {\bibinfo  {journal}
  {Phys. Lett. B}\ }\textbf {\bibinfo {volume} {28}},\ \bibinfo {pages} {493}
  (\bibinfo {year} {1969})}\BibitemShut {NoStop}%
\bibitem [{\citenamefont {Casas}\ and\ \citenamefont
  {Ibarra}(2001)}]{Casas:2001sr}%
  \BibitemOpen
  \bibfield  {author} {\bibinfo {author} {\bibfnamefont {J.~A.}\ \bibnamefont
  {Casas}}\ and\ \bibinfo {author} {\bibfnamefont {A.}~\bibnamefont {Ibarra}},\
  }\bibfield  {title} {\enquote {\bibinfo {title} {{Oscillating neutrinos and
  $\mu \to e, \gamma$}},}\ }\href {\doibase 10.1016/S0550-3213(01)00475-8}
  {\bibfield  {journal} {\bibinfo  {journal} {Nucl. Phys. B}\ }\textbf
  {\bibinfo {volume} {618}},\ \bibinfo {pages} {171--204} (\bibinfo {year}
  {2001})},\ \Eprint {http://arxiv.org/abs/hep-ph/0103065}
  {arXiv:hep-ph/0103065} \BibitemShut {NoStop}%
\bibitem [{\citenamefont {Capozzi}\ \emph {et~al.}(2021)\citenamefont
  {Capozzi}, \citenamefont {Di~Valentino}, \citenamefont {Lisi}, \citenamefont
  {Marrone}, \citenamefont {Melchiorri},\ and\ \citenamefont
  {Palazzo}}]{Capozzi:2021fjo}%
  \BibitemOpen
  \bibfield  {author} {\bibinfo {author} {\bibfnamefont {Francesco}\
  \bibnamefont {Capozzi}}, \bibinfo {author} {\bibfnamefont {Eleonora}\
  \bibnamefont {Di~Valentino}}, \bibinfo {author} {\bibfnamefont {Eligio}\
  \bibnamefont {Lisi}}, \bibinfo {author} {\bibfnamefont {Antonio}\
  \bibnamefont {Marrone}}, \bibinfo {author} {\bibfnamefont {Alessandro}\
  \bibnamefont {Melchiorri}}, \ and\ \bibinfo {author} {\bibfnamefont
  {Antonio}\ \bibnamefont {Palazzo}},\ }\bibfield  {title} {\enquote {\bibinfo
  {title} {{Unfinished fabric of the three neutrino paradigm}},}\ }\href
  {\doibase 10.1103/PhysRevD.104.083031} {\bibfield  {journal} {\bibinfo
  {journal} {Phys. Rev. D}\ }\textbf {\bibinfo {volume} {104}},\ \bibinfo
  {pages} {083031} (\bibinfo {year} {2021})},\ \Eprint
  {http://arxiv.org/abs/2107.00532} {arXiv:2107.00532 [hep-ph]} \BibitemShut
  {NoStop}%
\bibitem [{\citenamefont {Esteban}\ \emph {et~al.}(2020)\citenamefont
  {Esteban}, \citenamefont {Gonzalez-Garcia}, \citenamefont {Maltoni},
  \citenamefont {Schwetz},\ and\ \citenamefont {Zhou}}]{Esteban:2020cvm}%
  \BibitemOpen
  \bibfield  {author} {\bibinfo {author} {\bibfnamefont {Ivan}\ \bibnamefont
  {Esteban}}, \bibinfo {author} {\bibfnamefont {M.~C.}\ \bibnamefont
  {Gonzalez-Garcia}}, \bibinfo {author} {\bibfnamefont {Michele}\ \bibnamefont
  {Maltoni}}, \bibinfo {author} {\bibfnamefont {Thomas}\ \bibnamefont
  {Schwetz}}, \ and\ \bibinfo {author} {\bibfnamefont {Albert}\ \bibnamefont
  {Zhou}},\ }\bibfield  {title} {\enquote {\bibinfo {title} {{The fate of
  hints: updated global analysis of three-flavor neutrino oscillations}},}\
  }\href {\doibase 10.1007/JHEP09(2020)178} {\bibfield  {journal} {\bibinfo
  {journal} {JHEP}\ }\textbf {\bibinfo {volume} {09}},\ \bibinfo {pages} {178}
  (\bibinfo {year} {2020})},\ \Eprint {http://arxiv.org/abs/2007.14792}
  {arXiv:2007.14792 [hep-ph]} \BibitemShut {NoStop}%
\bibitem [{\citenamefont {de~Salas}\ \emph {et~al.}(2021)\citenamefont
  {de~Salas}, \citenamefont {Forero}, \citenamefont {Gariazzo}, \citenamefont
  {Mart\'\i{}nez-Mirav\'e}, \citenamefont {Mena}, \citenamefont {Ternes},
  \citenamefont {T\'ortola},\ and\ \citenamefont {Valle}}]{deSalas:2020pgw}%
  \BibitemOpen
  \bibfield  {author} {\bibinfo {author} {\bibfnamefont {P.~F.}\ \bibnamefont
  {de~Salas}}, \bibinfo {author} {\bibfnamefont {D.~V.}\ \bibnamefont
  {Forero}}, \bibinfo {author} {\bibfnamefont {S.}~\bibnamefont {Gariazzo}},
  \bibinfo {author} {\bibfnamefont {P.}~\bibnamefont {Mart\'\i{}nez-Mirav\'e}},
  \bibinfo {author} {\bibfnamefont {O.}~\bibnamefont {Mena}}, \bibinfo {author}
  {\bibfnamefont {C.~A.}\ \bibnamefont {Ternes}}, \bibinfo {author}
  {\bibfnamefont {M.}~\bibnamefont {T\'ortola}}, \ and\ \bibinfo {author}
  {\bibfnamefont {J.~W.~F.}\ \bibnamefont {Valle}},\ }\bibfield  {title}
  {\enquote {\bibinfo {title} {{2020 global reassessment of the neutrino
  oscillation picture}},}\ }\href {\doibase 10.1007/JHEP02(2021)071} {\bibfield
   {journal} {\bibinfo  {journal} {JHEP}\ }\textbf {\bibinfo {volume} {02}},\
  \bibinfo {pages} {071} (\bibinfo {year} {2021})},\ \Eprint
  {http://arxiv.org/abs/2006.11237} {arXiv:2006.11237 [hep-ph]} \BibitemShut
  {NoStop}%
\bibitem [{\citenamefont {Hambye}\ \emph {et~al.}(2004)\citenamefont {Hambye},
  \citenamefont {Lin}, \citenamefont {Notari}, \citenamefont {Papucci},\ and\
  \citenamefont {Strumia}}]{Hambye:2003rt}%
  \BibitemOpen
  \bibfield  {author} {\bibinfo {author} {\bibfnamefont {Thomas}\ \bibnamefont
  {Hambye}}, \bibinfo {author} {\bibfnamefont {Yin}\ \bibnamefont {Lin}},
  \bibinfo {author} {\bibfnamefont {Alessio}\ \bibnamefont {Notari}}, \bibinfo
  {author} {\bibfnamefont {Michele}\ \bibnamefont {Papucci}}, \ and\ \bibinfo
  {author} {\bibfnamefont {Alessandro}\ \bibnamefont {Strumia}},\ }\bibfield
  {title} {\enquote {\bibinfo {title} {{Constraints on neutrino masses from
  leptogenesis models}},}\ }\href {\doibase 10.1016/j.nuclphysb.2004.06.027}
  {\bibfield  {journal} {\bibinfo  {journal} {Nucl. Phys. B}\ }\textbf
  {\bibinfo {volume} {695}},\ \bibinfo {pages} {169--191} (\bibinfo {year}
  {2004})},\ \Eprint {http://arxiv.org/abs/hep-ph/0312203}
  {arXiv:hep-ph/0312203} \BibitemShut {NoStop}%
\bibitem [{\citenamefont {Strumia}(2006)}]{Strumia:2006qk}%
  \BibitemOpen
  \bibfield  {author} {\bibinfo {author} {\bibfnamefont {Alessandro}\
  \bibnamefont {Strumia}},\ }\bibfield  {title} {\enquote {\bibinfo {title}
  {{Baryogenesis via leptogenesis}},}\ }in\ \href@noop {} {\emph {\bibinfo
  {booktitle} {{Les Houches Summer School on Theoretical Physics: Session 84:
  Particle Physics Beyond the Standard Model}}}}\ (\bibinfo {year} {2006})\
  pp.\ \bibinfo {pages} {655--680},\ \Eprint
  {http://arxiv.org/abs/hep-ph/0608347} {arXiv:hep-ph/0608347} \BibitemShut
  {NoStop}%
\bibitem [{\citenamefont {Buchmuller}\ \emph
  {et~al.}(2005{\natexlab{b}})\citenamefont {Buchmuller}, \citenamefont
  {Di~Bari},\ and\ \citenamefont {Plumacher}}]{Buchmuller:2004nz}%
  \BibitemOpen
  \bibfield  {author} {\bibinfo {author} {\bibfnamefont {W.}~\bibnamefont
  {Buchmuller}}, \bibinfo {author} {\bibfnamefont {P.}~\bibnamefont {Di~Bari}},
  \ and\ \bibinfo {author} {\bibfnamefont {M.}~\bibnamefont {Plumacher}},\
  }\bibfield  {title} {\enquote {\bibinfo {title} {{Leptogenesis for
  pedestrians}},}\ }\href {\doibase 10.1016/j.aop.2004.02.003} {\bibfield
  {journal} {\bibinfo  {journal} {Annals Phys.}\ }\textbf {\bibinfo {volume}
  {315}},\ \bibinfo {pages} {305--351} (\bibinfo {year}
  {2005}{\natexlab{b}})},\ \Eprint {http://arxiv.org/abs/hep-ph/0401240}
  {arXiv:hep-ph/0401240} \BibitemShut {NoStop}%
\bibitem [{\citenamefont {Nardi}\ \emph {et~al.}(2007)\citenamefont {Nardi},
  \citenamefont {Racker},\ and\ \citenamefont {Roulet}}]{Nardi:2007jp}%
  \BibitemOpen
  \bibfield  {author} {\bibinfo {author} {\bibfnamefont {Enrico}\ \bibnamefont
  {Nardi}}, \bibinfo {author} {\bibfnamefont {Juan}\ \bibnamefont {Racker}}, \
  and\ \bibinfo {author} {\bibfnamefont {Esteban}\ \bibnamefont {Roulet}},\
  }\bibfield  {title} {\enquote {\bibinfo {title} {{CP violation in
  scatterings, three body processes and the Boltzmann equations for
  leptogenesis}},}\ }\href {\doibase 10.1088/1126-6708/2007/09/090} {\bibfield
  {journal} {\bibinfo  {journal} {JHEP}\ }\textbf {\bibinfo {volume} {09}},\
  \bibinfo {pages} {090} (\bibinfo {year} {2007})},\ \Eprint
  {http://arxiv.org/abs/0707.0378} {arXiv:0707.0378 [hep-ph]} \BibitemShut
  {NoStop}%
\bibitem [{\citenamefont {D'Onofrio}\ \emph {et~al.}(2014)\citenamefont
  {D'Onofrio}, \citenamefont {Rummukainen},\ and\ \citenamefont
  {Tranberg}}]{DOnofrio:2014rug}%
  \BibitemOpen
  \bibfield  {author} {\bibinfo {author} {\bibfnamefont {Michela}\ \bibnamefont
  {D'Onofrio}}, \bibinfo {author} {\bibfnamefont {Kari}\ \bibnamefont
  {Rummukainen}}, \ and\ \bibinfo {author} {\bibfnamefont {Anders}\
  \bibnamefont {Tranberg}},\ }\bibfield  {title} {\enquote {\bibinfo {title}
  {{Sphaleron Rate in the Minimal Standard Model}},}\ }\href {\doibase
  10.1103/PhysRevLett.113.141602} {\bibfield  {journal} {\bibinfo  {journal}
  {Phys. Rev. Lett.}\ }\textbf {\bibinfo {volume} {113}},\ \bibinfo {pages}
  {141602} (\bibinfo {year} {2014})},\ \Eprint {http://arxiv.org/abs/1404.3565}
  {arXiv:1404.3565 [hep-ph]} \BibitemShut {NoStop}%
\end{thebibliography}%

\end{document}